%% file: main.tex
\pdfoutput=1
\documentclass[acmlarge]{acmart}

\input{Preamble/preamble}
\input{Preamble/macros}
\begin{document}

\title{\THEMIS: A Decentralized Privacy-Preserving Ad Platform with Reporting Integrity}

\author{Gonçalo Pestana}
\affiliation{%
  \institution{Brave Software}
  \country{UK}
}

\author{Iñigo Querejeta-Azurmendi}
\affiliation{%
  \institution{Universidad Carlos III Madrid / ITFI, CSIC.}
  \country{Spain}
}

\author{Panagiotis Papadopoulos}
\affiliation{%
  \institution{Telefonica Research}
  \country{Spain}
}

\author{Benjamin Livshits}
\affiliation{%
  \institution{Brave Software/Imperial College London}
  \country{UK}
}

\makeatletter
\let\@authorsaddresses\@empty
\makeatother

\settopmatter{printacmref=false}
\setcopyright{none}
\renewcommand\footnotetextcopyrightpermission[1]{}

\renewcommand{\emph}[1]{\textit{#1}}

\input{Sections/00_abstract}
\maketitle

\input{Sections/01_introduction}
\input{Sections/02_background}
\input{Sections/03_requirements}
\input{Sections/04a_design}
\input{Sections/04b_themis}
\input{Sections/05_implementation}
\input{Sections/06_evaluation}
\input{Sections/07_discussion}
\input{Sections/08_related}
\input{Sections/09_conclusions}

\normalem
\bibliographystyle{unsrt}
\bibliography{main}

\appendix
\input{Sections/10_appendix_smart_contract}
\end{document}

%% file: Preamble/preamble.tex
\usepackage{tablefootnote}
\usepackage{thmtools}
\usepackage{thm-restate}
\usepackage{etoolbox}
\usepackage{xcolor,colortbl}
\usepackage{bigstrut}
\usepackage{booktabs}
\usepackage{graphicx}
\usepackage{caption}
\usepackage[T1]{fontenc}
\usepackage[utf8]{inputenc}
\usepackage{listings}
\usepackage{xspace}
\usepackage{amsmath}
\usepackage{amsthm}

\makeatletter
\patchcmd{\section}
  {\leavevmode\vrule\@width\z@\@height\dimexpr\topskip+6.5\p@\relax\@depth\z@}
  {}
  {}{}
\makeatother

\usepackage[compact]{titlesec} 
\theoremstyle{definition}
\newtheorem{phase}{Phase}[subsection]

\usepackage{tikz}
\usepackage{enumitem}
\usepackage{pgfplots} 
\usepackage{pgfplotstable}
\usepackage{ulem} 
\usepackage{hyperref}

\pgfplotstableread{
x         y    y-max  y-min
Request  1.256 0.02 0.025
Fetch/Decrypt/Recover 0.469 0.015 0.011 
}{\client}

%% file: Preamble/macros.tex
\newcommand{\code}[1]{\textsf{#1}\xspace}

\newcommand{\point}[1]{\vspace{0.01in}\par\vspace{0.01in}\noindent{\textbf{#1:} }}

\newcommand{\nrads}{n_A}

\newcommand{\advs}{$Advs$\xspace}
\newcommand{\cp}{\texttt{Con.Part}\xspace}
\newcommand{\nrcp}{n} 
\newcommand{\threshold}{k}

\newcommand{\ad}{\code{Ad}}

\newcommand{\empirical}[1]{#1}

\newcommand{\ecenc}{\code{Encrypt}}

\newcommand{\pk}{\code{pk}}
\newcommand{\sk}{\code{sk}}

\newcommand{\signature}{\sigma}

\newcommand{\distpk}{\pk_{\code{T}}}
\newcommand{\distsk}{\sk_{\code{T}}}

\newcommand{\distski}[1]{\sk_{\code{T},#1}}

\newcommand{\encryptedad}{\code{Enc}.\ad}

\newcommand{\encryptedaggregatedresult}{\code{Aggr.Res}}
\newcommand{\encryptedaggregatedclicks}{\code{Aggr.Clicks}}
\newcommand{\decryptedaggr}{\code{Dec.\encryptedaggregatedresult}}
\newcommand{\result}{\code{Res}}
\newcommand{\aggrproofdec}{\Pi_{\result}}
\newcommand{\aggrclicks}{\code{Aggr.Clicks}}

\newcommand{\vrf}{\code{VRF}}
\newcommand{\vrfpk}{\vrf.\pk}
\newcommand{\vrfsk}{\vrf.\sk}

\newcommand{\vrfrandgen}{\vrf.\code{RandGen}}
\newcommand{\vrfrandom}{\vrf.\code{Rand}}
\newcommand{\vrfproof}{\Pi^{\vrf}}
\newcommand{\vrfseed}{\epsilon}
\newcommand{\vrfmax}{\code{MAX.DRAW}}

\newcommand{\encvec}{\code{EncVec}}
\newcommand{\encvecpublic}{\code{EncVec'}}

\newcommand{\signaturereward}{\code{SignReward}}

\newcommand{\policy}{\mathcal{P}}
\newcommand{\encpolicy}{\code{Enc }\mathcal{P}}
\newcommand{\seckey}{\mathcal{S}}
\newcommand{\encseckey}{\code{Enc }\mathcal{S}}
\newcommand{\valikey}{\pk_{\mathcal{V}}}

\newcommand{\addr}{\code{Addr}}
\newcommand{\payreq}{\mathcal{E}}

\newcommand{\cfpk}{\code{\cf.}\pk}

\newcommand{\adclicks}{\code{ac}}%
\newcommand{\nrad}{\code{n}}%

\newcommand{\THEMIS}{THEMIS\xspace}
\newcommand{\PoA}{\textsc{PoA}\xspace}
\newcommand{\PoW}{\textsc{PoW}\xspace}
\newcommand{\PSC}{\textsc{PSC}\xspace}
\newcommand{\FSC}{\textsc{FSC}\xspace}

\newcommand{\cf}{CF\xspace}
\newcommand{\cfs}{CFs\xspace}
\newcommand{\cm}{CM\xspace}
\newcommand{\sidechain}{sidechain\xspace}
\newcommand{\sidechains}{sidechains\xspace}

\newcommand{\adcatalog}{ad-catalog\xspace}

\newcommand{\eg}{{e.g.,}\xspace}
\newcommand{\etc}{{etc.}\xspace}

\newcommand{\ie}{{i.e.,}\xspace}

\newcommand{\enc}{\code{Enc}}
\newcommand{\dec}{\code{Dec}}
\newcommand{\sign}{\code{Sign}}
\newcommand{\dpoolsize}{L\xspace} 

\newcommand{\paymentsPerDay}{\empirical{1.7M}\xspace}
\newcommand{\batchsize}{\empirical{800}\xspace}
\newcommand{\paymentsPerMonth}{\empirical{50.9M}\xspace}
\newcommand{\usersScalability}{\empirical{51M}\xspace}
\newcommand{\usersScalabilityMulti}{\empirical{153M}\xspace}

\newcommand{\rewardrequest}{\empirical{4.69 seconds}\xspace}
\newcommand{\rewardverif}{\empirical{5.04 seconds}\xspace}

\newcommand{\partdec}{\code{Part.Dec}\xspace}
\newcommand{\decsharei}[1]{\code{Dec.Share}_{#1}}

\usepackage[framemethod=default]{mdframed}
\newenvironment{cvbox}[1][]{%
    \begin{mdframed}[%
        frametitle={\small#1},
        font=\footnotesize
    ]%
}{%
    \end{mdframed}
}

%% file: Sections/00_abstract.tex
\begin{abstract}
{Online advertising fuels the (seemingly) free internet. However, although users can access most of the web services free of charge, they pay a heavy cost on their privacy. They are forced to trust third parties and intermediaries, who not only collect behavioral data but also absorb great amounts of ad revenues. Consequently, more and more users opt out from advertising by resorting to ad blockers, thus costing publishers millions of dollars in lost ad revenues. Albeit there are various privacy-preserving advertising proposals (\eg Adnostic, Privad, Brave Ads) from both academia and industry, they all rely on centralized management that users have to blindly trust without being able to audit, while they also fail to guarantee the integrity of the performance analytics they provide to advertisers.
In this paper, we design and deploy \THEMIS, a novel, decentralized and privacy-by-design ad platform that requires zero trust by users. \THEMIS (i) provides auditability to its participants, (ii) rewards users for viewing ads, and (iii) allows advertisers to verify the performance and billing reports of their ad campaigns. By leveraging smart contracts and zero-knowledge schemes, we implement a prototype of \THEMIS and early performance evaluation results show that it can scale linearly on a multi-\sidechain setup while it supports more than~\usersScalability users on a single-\sidechain.}
\end{abstract}

%% file: Sections/01_introduction.tex
\section{Introduction}
\label{sec:intro}

Despite the various alternative monetization systems~\cite{paywalls,wikipediaFunds1,truth2018,10.1145/1963405.1963451}, there is no doubt that digital advertising is still the dominant way of funding the web.
However, digital advertising has fundamental flaws, including market fragmentation~\cite{frag1, frag2}, rampant fraud~\cite{fraud1, fraud2,adfraud,Zarras:2014:DAM:2663716.2663719,malvertising}, centralization around intermediaries (\eg Supply/Demand-side platforms, Ad exchanges, Data Management Platforms), who absorb half of spent advertising dollars~\cite{middlemenHalf}, and cause unprecedented privacy harm by extensively track users~\cite{trackersWWW2016,exclusiveCSync,englehardt2016online}. Additionally, advertising is increasingly being ignored or blocked by users: 
the average click-through-rate today is~2\% ~\cite{ctrOverall}, while at the same time~47\% of Internet users globally use ad-blockers~\cite{adblockreport2019}, thus costing publishers millions of dollars in ad revenues every year~\cite{adblockersCost}.

Academia and industry have responded to some of these challenges by designing novel ad platforms that emphasize end-user choice, privacy protections, fraud prevention, or performance improvements. Privad~\cite{privad}, Adnostic~\cite{toubiana2010adnostic} and Brave Ads~\cite{Brave2017} are among the three most prominent such proposals. 
An important novelty of the latter, browser-based advertising system (in operation since 2019) is that users are compensated for their attention while viewing ads in the form of a in the form of a cryptocurrency reward in their built-in browser wallet. As measured, 
this aspect increases user engagement achieving (14\% click-through-rate compared to the~2\% of traditional ad ecosystem)~\cite{ctrBrave}.

But although promising, yet these approaches have limitations that prevent them from being compelling replacements of the existing web advertising: they either do not scale well, or heavily rely on central authorities to process ad transactions. Additionally, these proposals lack auditability: parties need to blindly trust the ad network to exclusively determine how much advertisers will be charged, as well as the revenue share the publishers will get~\cite{7164916}. Malicious ad networks can overcharge advertisers or underpay publishers, while malicious advertisers can deny actual views/clicks and ask for refunds, since bills frequently cannot be adequately justified by the ad network (non-repudiation). In case of the user rewards schema of Brave Ads, the computation of the rewards per user and the billing for the advertisers is performed by a central entity, and nobody can verifies their validity.


To address these issues, we propose \THEMIS, the first decentralized, scalable, and privacy-preserving ad platform that provides auditability to its participants and (following Brave Ads paradigm) incentivizes users to interact with ads by rewarding them for their interactions with ads. As a result, in \THEMIS, users do not need to trust \emph{any} protocol participant (\eg to protect their privacy or to handle rewards). In addition, \THEMIS provides advertisers with verifiable and privacy-preserving campaign performance analytics. This way, advertisers can accurately learn \emph{how many users viewed their ads, but without learning which users}. 

To achieve the above, our system uses a \sidechain design pattern and smart contracts to eliminate the centralized management and brokerage of the current ad ecosystem. To reward users for their ad viewing and perform reward payouts without revealing their browsing patterns, \THEMIS leverages a partial homomorphic encryption scheme.
\THEMIS is privacy-preserving; it does not reveal any behavioral user data, while at the same time, it serves relevant ads to user's interests by matching ads with the user profile. Following other industry and academia proposals, the ad matching happens on device to preserve the user's privacy. \THEMIS leverages zero-knowledge schemes for its operations, so every actor in the system can cryptographically verify at any time that everybody follows the protocol as expected, without revealing critical user and advertiser data. Finally, by using a multiparty computation protocol, \THEMIS guarantees the computational integrity of the ad campaign performance analytics provided to advertisers.

\point{Contributions} In summary, we make the following contributions:
\begin {enumerate}[topsep=0pt]
\item We propose \THEMIS, a novel privacy-preserving advertising platform that rewards users for viewing ads. Contrary to existing proposals, our system is decentralized and leverages smart contracts to orchestrate reward calculation and payments between users and advertisers. This way, our platform avoids relying on a single trusted central authority.

\item To assess the feasibility of our approach, we implement a prototype of our system in Rust and Solidity. 
We provide the source code of \THEMIS publicly\footnote{Link to open-source code: \url{https://github.com/themis-ads/prototype}}. 

\item To evaluate the scalability and practicability of our system, we perform experiments on a single- and different multi-\sidechain setups. We see that \THEMIS can support reward payments of around~\usersScalability  users per month on a single \sidechain or~\usersScalabilityMulti users on a parallel setup of three \sidechains, showing that \THEMIS scales linearly with the number of \sidechains.
\end {enumerate}

%% file: Sections/02_background.tex
\section{Building Blocks}
\label{sec:blocks}

In this section, we provide the necessary background of the techniques and mechanisms used in \THEMIS.

\subsection{Proof of Authority Blockchains}
\label{subsec:poa}
\THEMIS relies on a blockchain with smart contract functionality to provide a decentralized ad platform. Smart contracts enable \THEMIS to perform the users' reward calculation and payments without needing to trust a central authority. Ethereum Mainnet is a popular Proof of Work (PoW), smart contract-based blockchain. However, due to its low transaction throughput, the high gas costs, and the overall poor scalability, in \THEMIS, we chose to use a \mbox{Proof-of-Authority}~(\PoA) blockchain instead. 

Consensus protocols constitute the basis of any distributed system. The decision of which consensus mechanism to use affects the properties, scalability and assumptions of the services built on top of the distributed system~\cite{consensus_sok}. A \PoA blockchain consists of a distributed ledger that relies on consensus achieved by a permissioned pool of validator nodes. \PoA validators can rely on fast consensus protocols such as IBFT/IBFT2.0~\cite{EIP650, ibft} and Clique~\cite{clique}, which result in faster minted blocks and thus \PoA can reach higher transaction throughput than traditional \PoW based blockchains. As opposed to traditional, permissionless blockchains (such as Bitcoin and Ethereum), the number of nodes participating in the consensus is relatively small and all nodes are authenticated. In \THEMIS, the set of validators may consist of publishers or foundations.


\point{Private transactions}
\label{sec:private_transactions} \THEMIS leverages private input transactions, which enable users to call smart contract functions with private inputs. More precisely, \THEMIS uses the private input transactions as defined by the Quorum \PoA \sidechain \cite{quorumpoa}. Providing private input functionality in smart contracts requires the inputs to be encrypted with all validator's public keys. 
By encrypting both inputs and outputs with validator's public keys, the parameters are private from readers of the public information while, at the same time, validator nodes can decrypt the values and run the smart contracts correctly in order to achieve consensus. For simplicity, we refer to the public keys of validators as a single one throughout the paper, and denote it with $\valikey$. Projects like Quorum~\cite{quorumpoa, quorumgo} and Hyperledger Besu~\cite{pegasys} are \mbox{Ethereum-based} distributed ledger implementations that implement private transaction inputs and outputs.

\subsection{Cryptographic Tools and Primitives}
\label{sec:cryptoTools}

\point {Confidentiality}
\THEMIS uses an additively homomorphic encryption scheme to calculate the reward payouts for each user, while keeping the user behavior (e.g. ad clicks) private. Given a \mbox{public-private} \mbox{key-pair} $(\pk, \sk)$, the encryption scheme is defined by three functions: first, the encrypt function, where given a public key and a message, it outputs a ciphertext, $C = \enc(\pk, M)$. Secondly, the decrypt function, that given a ciphertext and a private key, it outputs a decrypted message, $M = \dec(\sk, C)$. And finally, the signing function, where given a message and a secret key, it outputs a signature on the message, $S = \sign(\sk, M)$. The additive homomorphic property guarantees that the addition of two ciphertexts, $C_1 = \enc(\pk, M_1), C_2 = \enc(\pk, M_2)$ encrypted under the same key, results in the addition of the encryption of its messages, more precisely, $\enc(\pk, M_1) + \enc(\pk, M_2) = \enc(\pk, M_1 + M_2)$.
We run our experiments using ElGamal scheme over elliptic curves~\cite{elgamalcrypto}. 

\point{Integrity} 
To prove correctness of the reward request, users in \THEMIS leverage zero knowledge proofs (ZPKs)~\cite{Goldwasser:1985:KCI:22145.22178}. This tool allows an entity (the prover) to prove to a different entity (the verifier) that a certain statement is true over a private input. This proof does not disclose any other information from the input other than whether the statement is true or not. \THEMIS leverages ZKPs to offload computation on the client-side, while maintaining integrity and privacy. We denote proofs with the letter $\Pi$, and use $\Pi.\code{Verify}$ to denote verification of a proof. 

\point{Distribution of trust}
\THEMIS generates a \mbox{public-private} \mbox{key-pair} for each ad campaign, under which sensitive user information is encrypted. In order to distribute the trust between multiple participants, \THEMIS leverages a distributed key generation~(DKG) protocol. This allows a group of participants to distributively generate the \mbox{key-pair} $(\distpk, \distsk)$, where each participant has a share of the private key, $\distski i$, and no participant ever gains knowledge of the full private key, $\distsk$.
The resulting \mbox{key-pair} is a $\threshold - \nrcp$ threshold \mbox{key-pair}, which requires at least $\threshold$ out of the $\nrcp$ participants that distributively generated the key, to interact during the decryption procedure. We follow the protocol presented in~\cite{Gennaro2007} for the key generation as well as the decryption procedure. 

In order to choose this selected group of key generation participants in a distributed way, \THEMIS leverages verifiable random functions~(VRFs) \cite{micalivrf, irtf-cfrg-vrf-05}. In general, VRFs enable users to generate a random number and prove its randomness. In \THEMIS, we use VRFs to select a random pool of users and generate the distributed keys. Given a \mbox{public-private} \mbox{key-pair} $(\vrfpk, \vrfsk)$, VRFs are defined by one function: random number generation, which outputs a random number and a zero knowledge proof of correct generation $(\vrfrandom, \vrfproof) = \vrfrandgen(\vrfsk, \vrfseed)$, where $\vrfseed$ is a random seed.

\point{Confidential payments for \mbox{account-based} blockchains} 
\label{sec:conf_payments_poa}
Confidential payments on \mbox{account-based} blockchains allow transfers of assets between accounts without disclosing the amount of assets being transferred or the balance of the accounts. Additionally, the sender proves the correctness of the payment (\ie prove that there was no double spending) using a zero knowledge proof. Confidential payments have drawn a lot of interest in both academia~\cite{aztecpaper,zether, zcashpaper} and industry~\cite{zcash, anonymouszether} recently. We use AZTEC~\cite{aztecpaper} as our underlying private payment system.

%% file: Sections/03_requirements.tex
\section{Threat Model and Design Goals}
\label{sec:motivation}
In this section, we introduce the main actors and threat model of \THEMIS. In addition, we describe the design principles of \THEMIS and how existing systems compare with it.

\subsection{Main Actors}
\point{\PoA validator nodes} \THEMIS leverages a \PoA \sidechain that relies on consensus achieved by a pool of permissioned validator nodes. The role of the validators is to mine the blocks of the \sidechain. In order to achieve this, each validator needs to evaluate the smart contract instructions against the user's inputs and global state, achieve consensus among the consortium on what is the next stage of the blockchain and mine new blocks. To preserve independence and the zero-trust requirement of the \sidechain, in \THEMIS, validator nodes are maintained by  non-colluding, independent third parties (\eg the Electronic Frontier Foundation (EFF), or a non-profit trade foundation of applications),  similar to existing volunteer networks like Tor, Gnutella or distributed VPNs~\cite{hola,varvello2019vpn0}. 

\point{Campaign facilitator (\cf)} 
The \cf is an entity authorized by the \PoA consortium that helps running the \THEMIS protocol. A \cf interacts with advertisers to agree on an ad policy of their preference and deploys the smart contracts in the \PoA \sidechain. In addition, the \cf is responsible for performing the confidential and verifiable payments to the users. The \cf has the role of a facilitator; \THEMIS requires an honest run of the protocol by the \cf for completeness, but not for verifiability. The system can detect when the \cf misbehaves, which provides the incentives for \cf to behave as expected by the protocol.

\point{Advertisers} 
The advertisers agree with the \cf the policies for each ad campaign they want to launch in \THEMIS. They receive an anonymized feedback for the performance of their campaigns. Advertisers can verify the validity of the reporting. In addition, advertisers can interact with the \PoA chain to verify that the amount charged for running campaigns corresponds to valid user interactions with campaign ads.

\point{Users} 
The users interact with the ads through an advertising platform, which selects and distributes the ads. Users interact with the \PoA \sidechain so their rewards are computed and paid. Greedy users may try to abuse the system and claim higher rewards. While this can be solved by using a daily cap, in the rest of the paper, we assume users running genuine software and following the protocol as described. In \THEMIS, users may also participate in a consensus pool where they interact with other users in a peer-to-peer way. We refer to them as consensus participants~(\cp).

\subsection{Threat Model}
\label{subsec:threat model}
In \THEMIS, we assume computationally bounded adversaries capable of (i) snooping communications, (ii) performing replay attacks, or (iii) cheating by not following the protocol.

One such adversary may act as a \cf aiming to collect more processing fees than agreed at the cost of user rewards or advertiser refunds. 
Another adversary may attempt to breach the user privacy and snoop their ad interactions. Such information could reveal interests, political/sexual/religious preferences,  that can be later sold or used beyond the control of the user~\cite{radioshakData,brokserSell,toysmartData}. Such an adversary may control at most $\threshold$ of the $\nrcp$ (see Section~\ref{sec:cryptoTools}) randomly selected users that are part of the consensus pool. This adversary is an \emph{adaptive} adversary, meaning that it can decide which participants to corrupt based on prior observations (\eg once the consensus pool has been selected). We require that $\threshold < \nrcp / 2$, which is optimal in the threshold encryption scenario~\cite{Gennaro2007}. Given that the consensus participants are chosen randomly, this means that, for ad interaction privacy, we assume that at most half of the consensus participants are malicious. 
Other adversaries may try to break the confidentiality of the advertisers' agreed ad policies and disclose rewarding strategies to competitors. We assume that such an adversary may act as a user and/or advertiser in the protocol, but cannot control the campaign facilitators or the \PoA validators. 
%
%

\point{Out-of-scope Attacks} 
We acknowledge that client-side fraud, together with malvertising and brand safety are  important issues of the ad industry. However, similar to the related work~\cite{toubiana2010adnostic,10.5555/646139.680791,10.1145/2462456.2464436,privad,green2016protocol}, in this paper, we do not claim to address all issues of digital advertising. There is an abundance of papers aiming to detect and prevent cases of client-side fraud (\ie bot clicks, click farms, sybil attacks), which can be also used in the context of THEMIS (\eg distributed user reputation systems~\cite{yang2019decentralized}, anomaly detection, bluff ads~\cite{haddadi2010fighting}, bio-metric systems~\cite{2019zksense}, client puzzles~\cite{4215910}, \etc).
We do, however, outline mechanisms that reduce the incentives for clients to cheat and to control the number of sybils that an adversary is able to run in the system. We discuss more about those strategies in Section~\ref{clientside_fraud}. 

\subsection{Goals and Comparison with Alternatives}
\label{subsec:reqs}
The goal of this paper is to design (i) a decentralized and (ii) trustless ad platform that (iii) is private-by-design while, at the same time, (iv) rewards users for the attention they give while viewing ads and (v) provides metrics for the ad campaigns of the advertisers. The key system properties we focus on while designing \THEMIS, include privacy, decentralization and auditability, and scalability:

\begin{enumerate}[leftmargin=0.5cm]
\item \textbf{Privacy.} In the context of a sustainable ad ecosystem, we define privacy as the ability for users and advertisers to use \THEMIS without disclosing any critical information:
\begin{enumerate}[leftmargin=0.5cm]
\item For the user, privacy means being able to interact (\ie view, click) with ads without revealing their interests/preferences to any third party. 
    In \THEMIS, we preserve the privacy of the user not only when they are interacting with ads but also when they claim the corresponding rewards for these ads. 

\item For the advertisers, privacy means that they are able to setup ad campaigns without revealing any policies (\ie what is the reward of each of their ads) to the prying eyes of their competitors. \THEMIS keeps the ad policies confidential throughout the whole process, while it enables users to provably claim rewards based on the agreed ad policies.
\end{enumerate}

\item \textbf{Decentralization and auditability.}
Existing ad platforms~\cite{Brave2017,privad,toubiana2010adnostic} require a single central authority to manage and orchestrate the execution of their protocols. Both privacy and billing is dependent on the correct behaviour of the single authority. 

However, as pointed out in~\cite{7164916}, what if this --- considered as \emph{trusted}~---~entity censors users by denying or transferring incorrect amount of rewards? What if it attempts to charge advertisers more than what they should pay based on users' ad interactions? What if the advertising policies are not applied as agreed with the advertisers when setting up ad campaigns?

One of the primary goals of our system is to be decentralized and require no trust from users. To achieve this, \THEMIS leverages a Proof-of-Authority (\PoA) blockchain with smart contract functionality. To provide auditability, \THEMIS leverages zero-knowledge proofs to ensure the correctness and validity of both billing and reporting thus allowing all actors to verify the authenticity of the statements and the performed operations.

\item \textbf{Scalability.}
Ad platforms need to be able to scale seamlessly and serve millions of users. However, important proposed systems fail to achieve this~\cite{privad,toubiana2010adnostic}. In this paper, we consider scalability an important aspect affecting the practicability of the system. \THEMIS needs to be able to serve ads in a privacy preserving way to millions of users and to finalize the payments related to user's ad interactions in a timely and resource efficient manner.

\end{enumerate}


%% file: Sections/04a_design.tex
\section{System Overview}
\label{sec:system}
In this section, we describe \THEMIS in detail. We begin with a straw-man approach to describe the basic principles of the system. We build on this straw-man approach and, \mbox{step-by-step}, we introduce the decentralized and trustless ad platform. For presentation purposes, in the rest of this paper, we assume that users interact with \THEMIS through a web browser, although users may interact with it through a mobile app in the exact same way. For the construction of \THEMIS, we assume the existence of a \mbox{privacy-preserving} ad personalization and the need for incentivizing users for interacting with ads. These requirements are currently being used by Brave Ads~\cite{Brave2017}, in continuous operation since 2019 and serving millions of users per month~\cite{BraveAds}. 

In \THEMIS, each viewed/clicked ad yields a reward, which can be fiat money, crypto-coins, coupons, \etc. Different ads may provide different amount of reward to the users.
Users \emph{claim} the rewards periodically (\eg every 2 days, every week or every month). Users request their rewards for the ads they viewed and interacted with. 

\begin{figure}[t]
    \centering
    \includegraphics[width=.45\columnwidth]{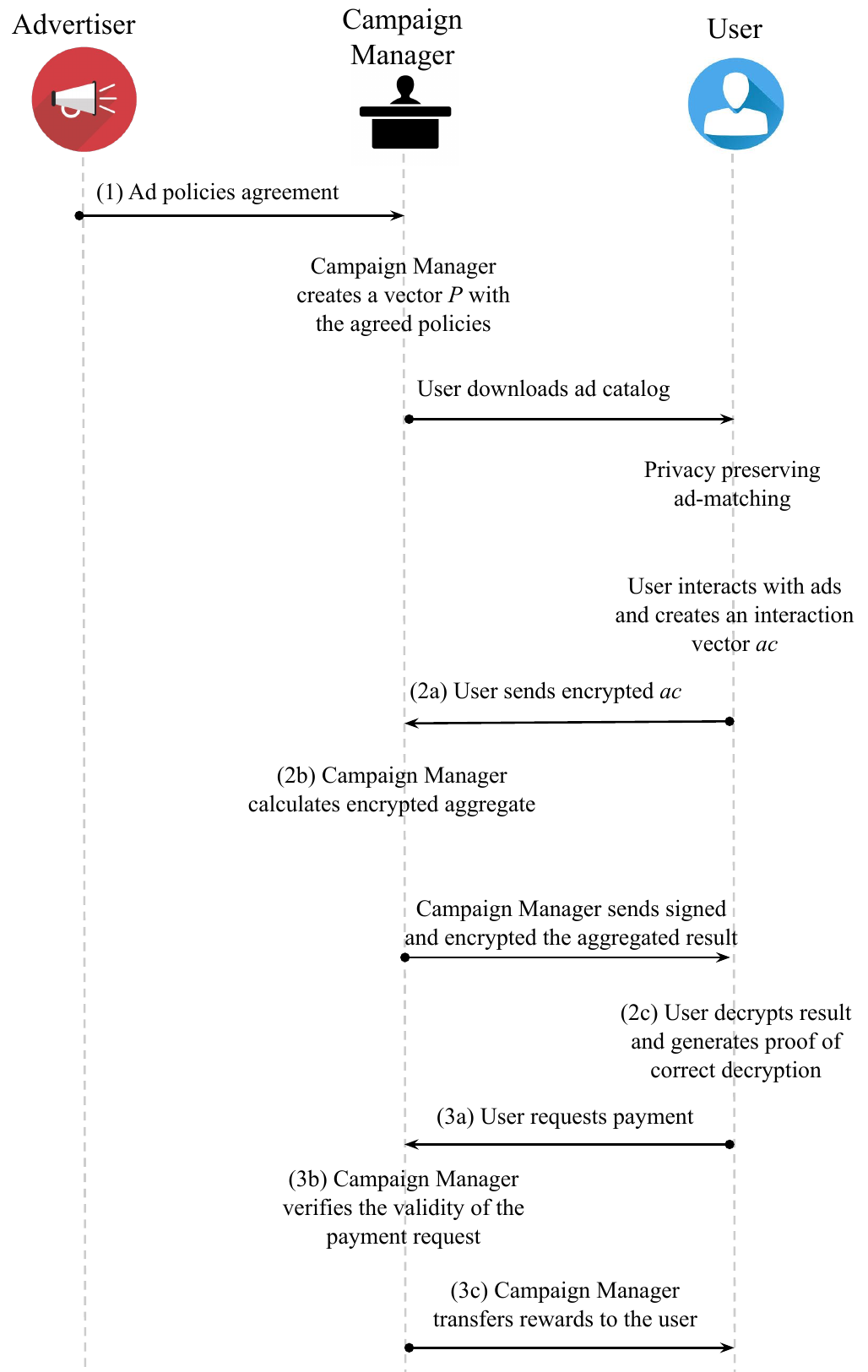}\vspace{-0.3cm}
    \caption{High-level overview of the user rewards claiming procedure of our straw-man approach. Advertisers can set how much they reward each ad click without disclosing that to competitors. The user can claim rewards without exposing which ads they interacted with.}
    \label{fig:diagram-strawman}
\end{figure}

\subsection{A Straw-man Approach}
\label{sec:strawman}
Our straw-man approach is the first step towards a privacy-preserving and trustless online advertising system. Our goal at this stage is to provide a mechanism for advertisers to create ad campaigns and to be correctly charged when their respective ads are delivered to users. In addition, the straw-main system aims at keeping track of the ads viewed by users, so that (i) advertisers can have feedback about the performance of the ad campaigns and (ii) users can be rewarded for interacting with ads. All these goals should be achieved while preserving ad policy privacy and user privacy.

We assume three different roles in the straw-man version of \THEMIS: (i) the users, (ii) the advertisers, and (iii) an ad campaigns manager (\cm). The users are incentivised to view and interact with ads created by the advertisers. The \cm is responsible (a) for orchestrating the protocol, (b) for handling the ad views reporting and finally (c) for calculating the rewards that need to be paid to users according to the policies defined by the advertisers.

Note that the straw-man version of \THEMIS relies on an ad campaign manager, which is a single central authority required for orchestrating the protocol. In addition, users and advertisers must trust the \cm. We use this simplified version to lay out the fundamental principles of the protocol. We present the improved -- and decentralized -- version of the protocol in Section~\ref{sec:demis}. In Figure~\ref{fig:diagram-strawman}, we present an overview of the reward and claiming procedure of the \THEMIS straw-man. 

\vspace{-0.15cm}\begin{phase}{\textbf{Defining Ad Rewards:}}
\label{phase:def-ads-strawman}
In order for an advertiser to include their ads in an ad campaign running on \THEMIS, they first need to agree with the \cm on the policies of the given campaign. An ad policy consists of the rewards a user should earn per ad visualization and engagement (step~1 in Figure~\ref{fig:diagram-strawman}). 

The \cm encodes the ad policies from multiple advertisers as a vector $\policy$, where each value corresponds to the amount of tokens that an ad yields when viewed/clicked (\eg Ad 1: 4 coins, Ad 2: 20 coins, Ad 3: 12 coins). The indices used in $\policy$ are aligned with the ones of the \adcatalog, which keeps the metadata of the ad to render on the user device.

For the sake of simplicity, throughout this section, we consider one advertiser who participates in our ad platform and runs multiple ad campaigns. In a real world scenario many advertisers can participate and run multiple ad campaigns simultaneously. We also consider ``agreed policies'' as the amount of coins an ad provides as reward for a click by a user.\vspace{-0.15cm}
\end{phase}
\begin{phase}{\textbf{Claiming Ad Rewards:}}
\label{phase:claim-rewards-strawman}
The user locally creates an \emph{interaction vector}, which encodes information about the number of times each ad of the \adcatalog was viewed/clicked (\eg Ad 1: was viewed 3 times, Ad 2: was viewed 0 times, Ad 3: was viewed 2 times). We refer to this vector as \adclicks. 

In every payout period, the user encrypts the state of the interaction vector. They generate a new ephemeral key pair $(\pk, \sk)$, to ensure the unlinkability of the payout requests and by using this key, they encrypt each entry of \adclicks and send the result, \encvec, to the \cm (step 2a in Figure~\ref{fig:diagram-strawman}).
The \cm cannot decrypt the received vector and thus cannot learn the user's ad interactions (and consequently their interests). Instead, they leverage the additive homomorphic property of the underlying encryption scheme (as described in Section~\ref{sec:cryptoTools}) to calculate the sum of all payouts based on the interactions encoded in \encvec (step 2b in Figure~\ref{fig:diagram-strawman}). 
We refer to the resulting ciphertext as \encryptedaggregatedresult.
Then, the \cm signs the computed aggregate result, \signaturereward, and sends the 2-tuple $(\encryptedaggregatedresult, \signaturereward)$ back to the user.

Upon receiving this tuple, (step 2c in Figure~\ref{fig:diagram-strawman}), 
the user verifies the signature of the result. If the signature is invalid, the user repeats the request to the \cm. If the signature is valid, the user proceeds with decrypting the result. The decrypted result, $\decryptedaggr$, is the final amount of rewards the user should receive for the interactions with ads encoded in the \encvec.
As a final step, the user proves the correctness of the performed decryption by creating a zero knowledge proof of correct decryption, $\aggrproofdec$.\vspace{-0.15cm}
\end{phase}

\begin{phase}{\textbf{Payment Request:}}
\label{phase:payment-request-strawman}
Finally, the user generates the payment request and sends it to the \cm. The request consists of $\left(\decryptedaggr, \encryptedaggregatedresult, \signaturereward, \aggrproofdec \right)$ (step 3a in Figure~\ref{fig:diagram-strawman})
As a next step, (step 3b in Figure~\ref{fig:diagram-strawman}), 
the \cm verifies that the payment request is valid. To this end it checks the signature and the proof of correctness. If it validates correctly, it proceeds with transferring the proper amount (equal to $\decryptedaggr$) of rewards to the user.\vspace{-0.15cm}
\end{phase}

\point{In summary} 
The straw-man \THEMIS guarantees that:
\begin{enumerate}[leftmargin=0.5cm,label=\Alph*.]
    \item The user receives the rewards they earned by interacting with ads. This happens without requiring the user to disclose to any party their ad interactions. 
    \item \cm is able to correctly apply the pricing policy of each ad without disclosing any information regarding the ad policies to users or potential competitors of the advertisers.
\end{enumerate}

%% file: Sections/04b_themis.tex
\subsection{\THEMIS: A Decentralized Ad Platform}
\label{sec:demis} 
The centralization and the lack of auditability of straw-man \THEMIS creates significant limitations with respect to the goals and threat model described in Section~\ref{sec:motivation}:
\begin{itemize}[leftmargin=0.5cm]
  \item Advertisers need to \emph{blindly trust} the \cm with the full custody of the rewards budget set for each ad campaign.
  \item Users and advertisers \emph{have to trust} that the \cm respects the agreed policies during payouts and transfers the correct amount of rewards 
  \item Advertisers \emph{do not} receive performance analytics of their ads (e.g. how many times an ad was viewed/clicked). Moreover, not even the \cm is able to retrieve such information.
\end{itemize}

As a result, similarly to existing approaches~\cite{privad,toubiana2010adnostic}, the entire protocol relies on the trustworthiness of a the single central authority. Moreover, users and advertisers do not have any mechanism to verify that the protocol runs as expected.

To address these issues, \THEMIS leverages a distributed \PoA ledger where business and payment logic are orchestrated by smart contracts. All participants of \THEMIS can verify that everyone runs the protocol correctly, thus requiring zero trust from any player regarding verifiability. In particular, we define two smart contracts (See Appendix~\ref{sec:appendix-smart-contracts} for full details of the smart contracts
structure): 
\begin {enumerate}[label=\Alph*., leftmargin=0.5cm]
    \item \emph{The  Policy Smart Contract}~(\PSC), which is responsible for the billing of users' rewards and validating the payment requests. Furthermore, it is in this smart contract that $\encpolicy$ is stored. 
    \item \emph{The Fund Smart Contract}~(\FSC), which receives and escrows the funds needed to run the campaign. The \FSC is responsible for releasing (i) the funds needed for settling valid payment requests, (ii) refunds for advertisers, and (iii) the processing fees for paying participants who help running the protocol.
\end{enumerate}
In \THEMIS, instead of the central trusted authority of the \cm, we introduce the role of a 
\emph{Campaigns Facilitator}~(\cf). The responsibilities of the \cf are to 
(i)~negotiate the policies (\eg rewards per ad, impressions per ad) of the advertisers;
(ii)~deploy smart contracts in the \PoA ledger; and, lastly, (iii)~handle the on-chain reward payments. Our system ensures that everybody can audit and verify the behaviour of \cfs, so advertisers can pick the \cf they prefer to collaborate with, based on their reputation. The \cf is incentivized to perform the tasks required to facilitate the \adcatalog, by receiving \emph{processing fees} from advertisers.

Finally, to provide reports of performance to advertisers regarding their ad campaigns, \THEMIS incentivises users to perform a multiparty protocol to compute ad interaction analytics in a privacy preserving manner. These participating users are referred to as the \emph{Consensus Pool}.


\begin{figure}[t]
    \includegraphics[width=.45\columnwidth]{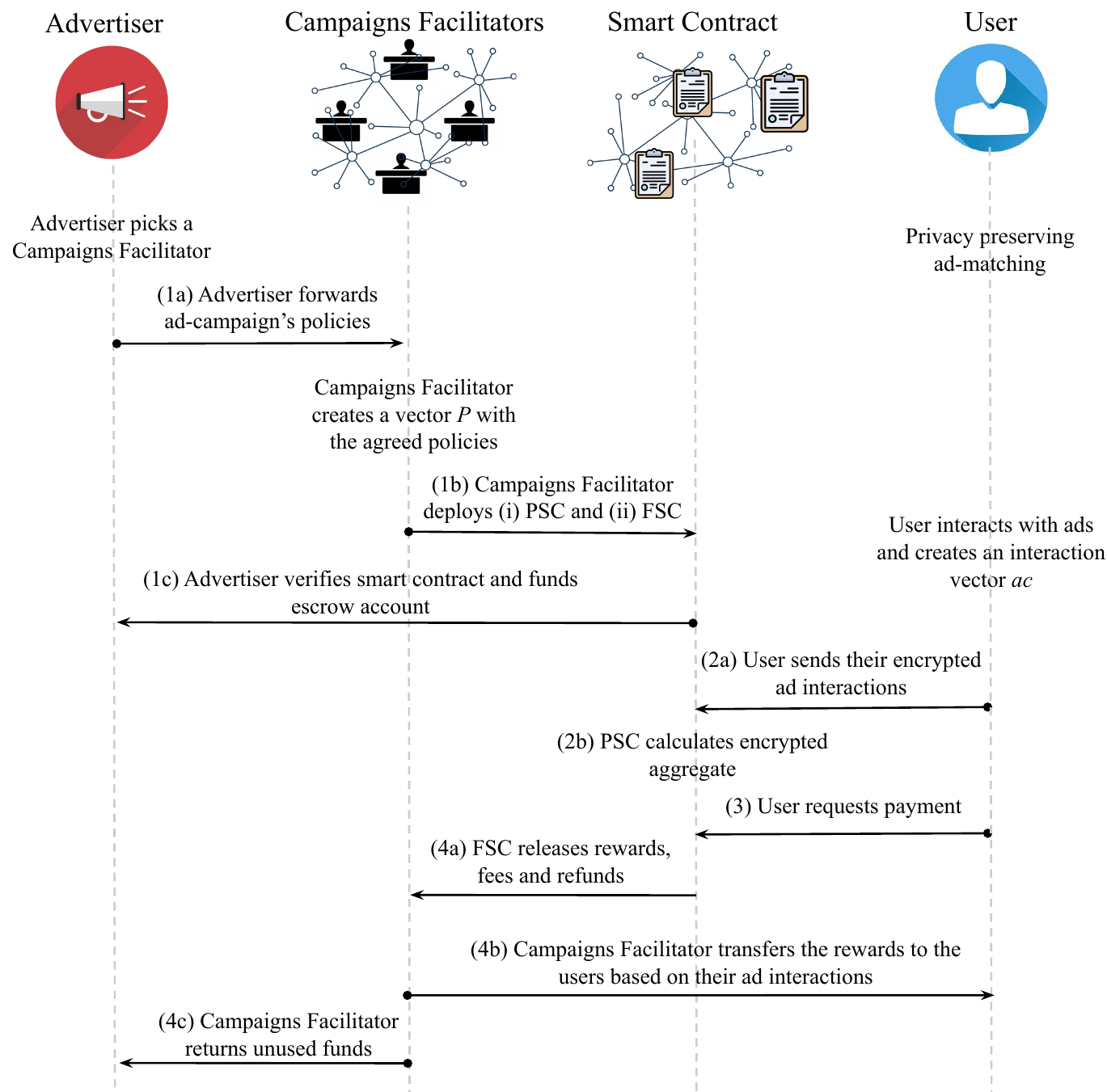}\vspace{-0.3cm}
    \caption{High-level overview of the user rewards claiming procedure in \THEMIS. This operation of \THEMIS consists of~4 different phases: (1)~Definition of Ad Rewards and set policies, (2)~Ad Reward claiming, (3)~Payment Request, (4)~Settlement of user payments and advertiser refunds.}
    \label{fig:diagram-demis}
\end{figure}

\vspace{-0.15cm}\begin{phase}{\textbf{Defining ad rewards:}}
\label{phase:def-ads-themis}
In Figure~\ref{fig:diagram-demis}, we present a high-level overview of the reward claiming procedure of \THEMIS. Similar to the straw-man approach presented in Section~\ref{sec:strawman}, in order for an advertiser to include their ad campaign in the next \adcatalog facilitated by the \cf of their preference, they need to transmit their policies (\eg reward per ad) to the \cf (step 1a in Figure~\ref{fig:diagram-demis}). In order to achieve that, each advertiser exchanges a symmetric key\footnote{For the creation of this key, they follow the Diffie-Hellman key exchange protocol~\cite{DHKE}} for each ad campaign $\seckey_i$ with the \cf. 
It then encrypts the corresponding ad campaign and sends it together with the ad data and metadata to the \cf. On their end, the \cf 
(i)~decrypts and checks the policies are as previously agreed with the advertiser; (ii)~merges the encrypted policies of the different advertisers into the encrypted policy vector, $\encpolicy$;  and 
(iii)~deploys both the \PSC and \FSC smart contracts for this \adcatalog version (step~1b in Figure~\ref{fig:diagram-demis}).
In addition, the \cf (iv) creates a vector $\seckey$ with all the advertisers' secret keys $\seckey_i$: 
\begin{equation}
    \seckey = \left[\seckey_1, \seckey_2 , \ldots, \seckey_{\nrads} \right] 
\end{equation}
and (v) generates a vector, $\encseckey$, that includes each of the elements in $\seckey$, encrypted with the public key ($\valikey$) of the \PoA validator nodes: 
\begin{equation}
    \encseckey =  \left[ \code{Enc}(\valikey, \seckey_1), \ldots, \code{Enc}(\valikey, \seckey_{\nrads}) \right]
\end{equation} 
Then, the \cf (vi) stores $\encseckey$ in \PSC to allow the  \PoA validators to decrypt and apply the corresponding policies on users ad interaction vectors. The process of encrypting the policy vector with symmetric keys, and then encrypting the symmetric keys with the validators public key is known as Hybrid Public Key Encryption~\cite{hpke}.

Once \PSC is deployed, the advertisers must verify if $\encpolicy$ encodes the policies agreed with \cf (step 1c in Figure~\ref{fig:diagram-demis}). More specifically:
\begin{enumerate}[leftmargin=0.5cm, label=\Alph*.]
    \item First, the advertisers fetch $\encpolicy$ vector from the public storage of the  \PSC and decrypt the policy, $\encpolicy[i]$, using the corresponding symmetric key, $\seckey_i$, and verify it is the agreed value.

    \item Second, they fetch the escrow account address from \FSC and transfer funds to the escrow account. The amount of funds needed is determined by the number of impressions they want per ad as defined in the agreed ad policy $\policy[i]$, and the processing fees to pay to the \cf. Once the campaign is over, the advertisers may get a refund based to the final number of impressions viewed/clicked by users. By staking the campaign's funds, the advertiser is implicitly validating the deployed ad policies.
\end{enumerate}
Once the \FSC verifies that advertisers have transferred to the escrow account the correct amount of funds, the campaign is considered verified and initialized.\vspace{-0.15cm}
\end{phase}
\begin{phase}{\textbf{Claiming ad rewards:}}
\label{phase:claimrewards-themis}
Similar to what is illustrated in Section~\ref{sec:strawman}, in order to claim their ad rewards, each user creates an ephemeral key pair $(\pk, \sk)$
and obtains the public threshold key $\distpk$ generated by the consensus pool (in Section~\ref{sec:aggrcount}, we describe in detail how the consensus pool is selected). By using these two keys, each user encrypts their ad interaction vector to generate two ciphertexts: 
(a)~the $\encvec$ that is used to claim ad rewards and (b)~the \encvecpublic that is used for the advertisers reporting. 

Contrary to our centralized straw-man approach, in \THEMIS, the aggregate calculation is performed by the smart contract \PSC (as can be seen in step~2b in Figure~\ref{fig:diagram-demis}). Thus, the user calls a public endpoint on \PSC with both ciphertexts as input. To calculate the encrypted sum of the rewards the user can claim (step~2b in Figure~\ref{fig:diagram-demis}), a \PoA validator runs \PSC as follows: 
\begin{enumerate}[leftmargin=0.5cm]
\item It decrypts each policy $\policy[i]$ using $\encseckey$ (here, \THEMIS leverages the private input transactions, see Section~\ref{sec:private_transactions}). 
\item It leverages the additively-homomorphic property of the underlying encryption scheme of $\encvec$ to compute the encrypted result.
\item It stores the encrypted result, $\encryptedaggregatedresult$, in the smart contract public store\footnote{Given that the user's public key was used for encrypting $\encvec$ ciphertext, only them can decrypt $\encryptedaggregatedresult$ and retrieve $\adclicks$.}.
\end{enumerate}\vspace{-0.15cm}
\end{phase}

\begin{phase}{\textbf{Payment request:}}
\label{phase:paymentrequest-themis}
Once the \PSC calculats the aggregate result (step~3 in Figure~\ref{fig:diagram-demis}), the user generates a payment request, $\payreq$, that, if valid, is published in \FSC. More specifically, the user 
(i)~creates an ephemeral account to receive the reward payment (used only once per payment request) with address \addr. Then (ii), it fetches and decrypts $\encryptedaggregatedresult$ to get the decrypted reward, \decryptedaggr, and generates the proof of correct decryption, $\aggrproofdec$. This way, the user 
(iii)~generates the payment request which consists of the following~3-tuple: 
\begin{equation}
    \payreq = \left[\decryptedaggr,\aggrproofdec, \addr \right].
\end{equation}
Then, (iv) the user calls a public endpoint on \PSC with the $\payreq$ encrypted with the validators keys, $\enc(\valikey, \payreq)$ as input. The function then fetches the user's aggregate, $\encryptedaggregatedresult$, decrypts the request, and verifies the zero knowledge proof, $\aggrproofdec$. If the proof is valid, it stores $\addr$ in the $\FSC$ together with the amount to be payed to this address. This way, the $\FSC$ keeps a list of buffered user payments until marked as paid.\vspace{-0.15cm}
\end{phase}
%
%
%
\begin{phase}{\textbf{Payment settlement:}}
\label{phase:payment-settlement-themis}
The final step of the protocol consists on the settlement of the user payment and advertiser refund. Specifically, the settlement of the user rewards in \THEMIS needs to happen in a confidential way to preserve the privacy of the total of earned rewards. To achieve this, the \cf fetches the pending payments requests from \FSC, and calculates the total amount of funds required to settle all pending payments. 

Next, (step~4a in Figure~\ref{fig:diagram-demis}), the \cf calls a public function of \FSC requesting to transfer (to an operational account owned by \cf) a given amount of tokens needed to cover the payments. If the \cf misbehaves (by requesting an incorrect amount of tokens), it will be detected, and either advertisers or users will be able to prove its misbehaviour. Finally, \cf settles each of the pending reward payments by using a confidential payment scheme (see Section~\ref{sec:cryptoTools}). After finalizing the payments correctly (and if there are no complaints form either users or advertisers), the \cf receives from \FSC the processing fees.

In case of unused staked funds, the advertisers need to be refunded. To achieve this (step~4c in Figure~\ref{fig:diagram-demis}), \FSC utilizes the aggregate clicks per ad vector that the consensus pool has computed during the advertisers reporting (see Section~\ref{sec:aggrcount}). Based on this vector and the agreed rewards, the \FSC proceeds with returning to the advertisers the unused funds.\vspace{-0.15cm}
\end{phase}

\subsection{Privacy-Preserving Performance Analytics}
\label{sec:aggrcount}
In addition to incentivizing users to interact with ads, in order to make an ad-platform practical, the advertisers must be able to receive feedback about their ad campaign performance. Moreover, advertisers need to verify that the funds charged correspond with the number of times an ad was viewed/clicked by the users. Based on these statistics, advertisers get charged depending on the times their corresponding ad was clicked throughout the campaign. 

To achieve this, whenever a new version of the \adcatalog is online and retrieved from the users, a new threshold public key, $\distpk$, is generated. In order to generate such a key, a pool of multiple participating users, namely \emph{consensus pool}, is created. 
To avoid cases where there are not enough participants available online, in \THEMIS the participation in the consensus pool is incentivised\footnote{Users are incentivized to participate in this pool. Details on how to orchestrate the incentives are out of scope of this paper.}.
The consensus pool consists of a number of selected users that have  \mbox{opted-in} to be part of it, and a smart contract responsible of the registration process and orchestration of the pool. Any user can \mbox{opt-in} in the draw to become a consensus pool participant, and a random subset of all participating users is selected. 
Specifically, the smart contract keeps a time interval during which users who want to participate can register as participants for the draft. After that,
the smart contract utilizes an external oracle to select a random seed,
$\vrfseed$, which is used to generate random numbers.

Every registered user generates an ephemeral Verifiable Random Function key-pair,
    $(\vrfpk, \vrfsk)$,
and publishes the public key in the smart contract's public store.
Once the registration phase is closed, the smart contract calculates a threshold, $\vrfmax$, that will define the selected users
\begin{equation*}
    \vrfmax = \lfloor \frac{\nrcp}{\dpoolsize}\rfloor * p
\end{equation*}
where \dpoolsize is the size of the drawing pool (formed by all opted-in users), $\nrcp$ the expected number of participants in the distributed key generation, and $p$ an integer such that $\mathbf{Z}_p$ is the space of random numbers outputted by $\vrfrandgen(\cdot, \cdot)$.

Next, the participating users calculate their corresponding random number and a proof of correct generation using $\vrfseed$:
\begin{equation*}
    (\vrfrandom, \vrfproof) = \vrfrandgen(\vrfsk, \vrfseed).
\end{equation*}
We consider that all participants with $\vrfrandom < \vrfmax$ win the draft and may participate in the consensus pool. The consensus pool participants proceed to publish $(\vrfrandom, \vrfproof)$ in the smart contract. 
Next, the selected consensus pool participants run a distributed key generation (DKG) algorithm as defined in~\cite{dkg}. The result of running the DKG protocol is a consensus over the public key to use in the rest of the process, $\distpk$. In addition, each consensus pool participant owns a private key share $\distski i$. The distributed public key, $\distpk$, is published in \PSC to make it accessible to all users.

This key is used to encrypt a copy of the \adclicks vector (Step 2a in Figure~\ref{fig:diagram-demis}). Hence, in addition to the \encvec, each user also sends \encvecpublic to the \PSC, where:
\begin{equation*}
\encvecpublic = \left[\ecenc(\distpk, \nrad_1), \ldots, \ecenc(\distpk, \nrad_{\nrads}) \right]
\end{equation*}
At the end of the ad campaign, the consensus pool generates the analytics report for the advertisers, showing how many times users interacted with each specific ad during the campaign. In order to generate the report, the consensus pool merges the reported \encvecpublic of every user into a single vector. This vector consists of the total number of interactions that each ad received by all users during the campaign. In order to merge all \encvecpublic of the campaign, the consensus pool performs the homomorphic addition of all reported encrypted vectors. This is possible due to the fact that every user used the same key for the encryption.  
Then, the consensus pool proceeds with the decryption, and the proof of correctness. More precisely, each consensus participant $i$ partially decrypts $\aggrclicks$: 
\begin{equation*}
    \decsharei i = \partdec(\distski i, \encryptedaggregatedclicks)
\end{equation*}
and proves it did so correctly. Finally, they post the encrypted aggregates of ads, the decrypted shares, and the proofs, to the \FSC. 

As soon as the \FSC receives at least the threshold $\threshold$ of such tuples, it combines the partial decryptions to compute the full decryption of the aggregates. This allows the advertisers to verify that the protocol ran successfully. 
The advertisers can now check that the encrypted aggregates calculated by the consensus pool corresponds with the addition of all encryptions submitted by users. To do this, each advertiser first performs the homomorphic addition of $\encryptedaggregatedresult'$. Next, they verify the proofs of correct decryption for each of the received shares. Then, they fetch the full decryption of the aggregate, representing the number of clicks/interactions their ads received.
If all these verifications succeed, advertisers can use the number of clicks/interactions their ad received to verify that the refund received by the \FSC does correspond to the number of staked funds minus the spent funds during the campaign.

%% file: Sections/05_implementation.tex
\section{Implementation}
\label{sec:implementation}
To assess the feasibility of our system, we implemented a prototype\footnote{The source code is publicly available: \url{https://github.com/themis-ads/prototype}} with the core components of the \THEMIS protocol that are necessary to run Phase~\ref{phase:def-ads-themis} (defining ad rewards on the client-side), Phase~\ref{phase:claimrewards-themis} (claiming ad rewards) and Phase~\ref{phase:paymentrequest-themis} (issuing and verifying the payment request).

\point{Smart contracts implementation} We implemented the policy smart contract in Solidity~\cite{solidity}. The policy smart contract is responsible to calculate the users payouts over encrypted input and it runs on the \sidechain Ethereum virtual machine (EVM). In addition, the smart contract verifies the proof of correct decryption sent by the users.

The cryptographic computations required by the policy smart contract are built upon a \mbox{pre-compiled} smart contract~\cite{eip196}, which implements the addition and scalar multiplication over the \texttt{alt\_bn128} curve. By relying on a \mbox{pre-compiled} smart contract, we improve the performance of our homomorphic operations running on the EVM. 

\point{Client-side implementation} We implemented the client logic in Rust. The client implementation leverages the \texttt{web3-rust}~\cite{web3-rust} crate to interact with the smart contracts. In addition, we used both \texttt{curve25519-dalek}~\cite{curve25519dalek} and \texttt{elgamal\_ristretto}~\cite{elgamal-rust} crates to implement the underlying \mbox{public-key} cryptography and corresponding operations, required by users to generate payment requests in \THEMIS.


%% file: Sections/06_evaluation.tex
\section{Evaluation}
\label{sec:results}
In this section, we study the performance and scalability of our system. First we set out to explore the execution time of the \mbox{client-side}  operations of \THEMIS: (i) rewards claiming and (ii) payment request.
Then, we study the \mbox{end-to-end} execution time when multiple participants request payments in \THEMIS. Finally, we measure the overall scalability of our system and specifically, how many concurrent users claiming rewards it supports, on both single and  multi-\sidechain setups. 

\input{Sections/06a_experimental_setup}
\begin{table}[t]
    \centering\footnotesize
    \caption{Execution time of the client-side operations during reward claiming for different \adcatalog s sizes.}\vspace{-0.3cm}
    \begin{tabular}{lrr}
        \toprule
        \bf Ad-Catalog & \bf Interaction & \bf Request  \\
        \bf Size (ads) & \bf Encryption & \bf Generation \\
        \midrule
        64 & 0.027 sec & 0.136 sec\\
        128 & 0.054 sec & 0.303 sec\\
        256 & 0.105 sec & 0.706 sec \\
        \bottomrule
    \end{tabular}
    \label{fig:client-side}\vspace{-0.4cm}
\end{table}

\subsection{System performance}
\label{sec:system-performance}
We set out to explore the execution time of client requests while participating in the \THEMIS protocol (\ie rewards claims and payment requests) and the time it takes for the \cf and the \sidechain to process these requests. Then, we measure the \mbox{end-to-end} time it takes for a reward request to be processed in our system.


In the case of users, we measure the time it takes for a client to generate locally a rewards claiming request for different \adcatalog sizes: 64, 128 and 256 ads, and in Table~\ref{fig:client-side} we present the results. As described in Phase~\ref{phase:claimrewards-themis}, this operation includes: 
\begin{enumerate} [leftmargin=0.5cm, label=(\roman*), itemsep=-0.1cm, topsep=-0.1cm]
\item Interaction encryption: includes the encryption of the interaction array of the user, and
\item Request generation: includes decryption of the payment aggregate, generation of the proof of correct decryption and recovery of plaintext.
\end{enumerate}

As can be seen, the execution time to encrypt user interactions for an \adcatalog of ~256 ads is as low as 0.1~sec. Similarly, for the same \adcatalog size, the request generation procedure takes around~0.7~sec, proving that the client computations for reward claiming can be done on a commodity laptop or mobile device, without significant impact on the user experience. Apart from issuing reward claiming requests, a client also performs periodic rewards payment requests. However, such requests take place in relatively long intervals (\eg monthly) and therefore the latency imposed to the user is practically negligible.

For the settlement of the rewards payment requests (Phase~\ref{phase:payment-settlement-themis}), the \cf relies on a confidential transaction protocol to ensure the confidentiality and integrity of the payments. We used the AZTEC protocol (Section \ref{sec:conf_payments_poa}) for measuring the performance of confidential payments. AZTEC allows batching of payments into a single \mbox{on-chain} proof verification of 3 seconds. This makes the verification time constant, independently of the number of batched payments. Our results, presented in Table~\ref{fig:aztec-scalability}, show that the \cf can achieve around~\paymentsPerDay payments/day for a batch size of~\batchsize proof payments. This results in a total of~\paymentsPerMonth payments.

As a next step, we measure the time required for the \sidechain to process concurrent payment requests \mbox{end-to-end}. This includes the generation of the payment requests by the user, the network latency in the communication between the clients and the \sidechain, and the time it takes for the \sidechain to process the requests and mine the blocks.

In Figure~\ref{fig:themis-several-users}, we show  the results with respect to the different concurrent users claiming their rewards (10, 30, 60, 100 users) and different \adcatalog sizes (yellow bar: 64 ads, blue bar: 128 ads, green bar: 256 ads). In red, we show the time it takes for the user to decrypt the aggregate, perform the plaintext recovery locally and submit the decrypted aggregate and proof of correct decryption. As we can see, to claim and retrieve rewards even in the case of 100 concurrent users and a large \adcatalog size of 256 ads\footnote{An \adcatalog large enough handle all different ads delivered simultaneously in production systems currently in use (\ie Brave Ads~\cite{BRAVE})}:
\begin{enumerate}[leftmargin=0.5cm, label=\roman*., itemsep=-0.1cm, topsep=-0.1cm]
	\item it takes roughly \rewardrequest for the client to request the reward calculation and retrieve the encrypted aggregate from the smart contract.
	\item it takes an additional 0.35 sec for the client to decrypt the aggregate, perform the plaintext recovery locally and submit the decrypted aggregate and proof of correct decryption.
\end{enumerate}
\noindent So it takes an overall of \rewardverif to process and verify up to~100 concurrent reward payment requests of users.
 
 \begin{table}[t]
    \centering\footnotesize
    \caption{Execution time of proof generation for batches of 80, 200, 400, and 800 concurrent payments of \cf.}\vspace{-0.3cm}
    \begin{tabular}{lrr}
    \toprule
        \bf Batched proofs & \bf Proof generation & \bf Proof verification \\
    \midrule
        80  & 3.9 sec & 3 sec \\
        200 & 11.6 sec & 3 sec \\
        400 & 22.08 sec & 3 sec \\
        800 & 40.7 sec & 3 sec\\
        \bottomrule
    \end{tabular}
    \label{fig:aztec-scalability}
\end{table}
 
\subsection{System Scalability}
\label{sec:system-scalability}
One of the most important challenges of \mbox{privacy-preserving} ad platforms is scalability. In the case of \THEMIS this related to how easy the system can scale with the increasing number of clients that simultaneously claim their rewards.

\input{Figs/fig-several-users}
%
%
%

In Figure~\ref{fig:themis-several-users}, we saw that for an \adcatalog of 256 ads and 100 concurrent users performing a payment request, it takes around~5 seconds to complete 100 concurrent payment requests. This means, that under the same conditions, the \sidechain can process around~1.7M concurrent payment requests/day, which translates to a total of \usersScalability users/month\footnote{The specifications used by the validator nodes to achieve this throughput are outlined in Section~\ref{sec:experimental-setup}.}.

\point{Horizontal scaling}
The computations performed by \THEMIS smart contracts are highly parallelizable. However, the \mbox{one-threaded} event loop of the Ethereum Virtual Machine (EVM)\footnote{The EVM is the \mbox{run-time} virtual machine where the smart contract instructions are executed in each of the \mbox{validator's} machines.} does not support parallel and concurrent computations. Therefore, the EVM \mbox{run-time} becomes the scalability bottleneck. In our experiments, it has been shown that the EVM outages when more than about 100 concurrent user requests are handled concurrently.

To overcome this shortcoming, we expect \THEMIS to run on top of multiple parallel \sidechains. Each \sidechain keeps its own state and is responsible for one or more distinct \adcatalog. Due to this, we expect scalability improvements to grow linearly with the number of \sidechains deployed, enabling the system to scale to support millions of concurrent users per day. 
 
\input{Figs/fig-multi-sidechains}

In order to explore how horizontal scaling performs in \THEMIS, we deployed and run load experiments with two and three parallel \sidechains, each using the same settings as outlined in Section~\ref{sec:experimental-setup}. Note that, by running multiple parallel \sidechains, it is not required additional validators. Instead, each validator is required to run multiple nodes, each node part of one single \sidechain.

Figure~\ref{fig:multiple-side-chains} shows the number of users \THEMIS can process by running on multiple \sidechains. As seen, the number of users increases linearly with the number of \sidechains. Assuming a setup with three parallel \sidechains, \THEMIS can handle a total of~\usersScalabilityMulti users per month~(5.1M users a day). Figure~\ref{fig:multiple-side-chains} also shows the estimated growth of number of users when more \sidechains are deployed.

Given the homomorphic property of the encryption scheme, the consensus participants need to perform addition of all ciphertexts, and decrypt a single vector (rather than decrypting all vectors). We ran the experiments in a commodity laptop (the one used for client performance), and ciphertext addition takes 273 nanoseconds (single threaded), which does not affect the overall scalability, even for~\usersScalabilityMulti users. 

\subsection{Summary}
In summary, evaluation results show that \THEMIS scales linearly and seamlessly supports user bases of existing, in-production, centralised systems ~\cite{BRAVE}. Specifically, our system
can support payment requests of around~\usersScalability  users on a single \sidechain setup or ~\usersScalabilityMulti  users on a parallel 3-\sidechain setup. 

In addition, we see that the latency users need to sustain while using \THEMIS is negligible (less that~1~sec per request payment on commodity hardware) and we show that both the \cf and \sidechain validators can also rely on commodity hardware to participate in the network.

%% file: Sections/06a_experimental_setup.tex
\subsection{Experimental Setup}
\label{sec:experimental-setup}

To study the performance and scalability of \THEMIS, we run \mbox{client-side} and \mbox{end-to-end} measurements using the \THEMIS prototype. In this section we outline specifications over which we run our experiments. 

\point{Client specifications} The \mbox{client-side} experiments were performed on a commodity device. The device is a MacBook Pro Catalina~10.15.5, running a~2.4GHz \mbox{Qual-Core} Intel Core i5 with~16GB LPDDR3 memory.

\point{Campaign facilitator specifications} 
To study the resources necessary for campaign facilitators to participate in the network, we measured the resource and time overhead required to generate and prove the correctness of confidential payments (Phase~\ref{phase:payment-settlement-themis} of \THEMIS). In order to do so, we deployed an AWS ECS \texttt{t2.2xlarge} instance (8~vCPUs, 32~GB~RAM). 

\point{Sidechain deployment} In order to measure the performance and scalability of the \sidechain in the context of \THEMIS, we used the Mjölnir tool~\cite{mjolnir} to deploy a Quorum~\cite{quorum} \sidechain in a \mbox{production-like} environment. We deployed a 4x~Quorum \sidechain on AWS, each node running on an AWS EC2 \texttt{t2.xlarge} instance (4~vCPUs, 16~GB~RAM). All nodes are deployed in the same AWS region and part of the same subnet. For the purpose of the measurements, the network communication is considered negligible. This setup can be easily reproduced in production by setting up peering connections among different AWS Virtual Private Clouds for each of the validator organizations. The consensus protocol used by the \sidechain is the Istanbul Byzantine Fault Tolerant (IBFT) consensus protocol~\cite{EIP650}.

\point{Concurrent users} A \mbox{production-like} environment requires multiple clients requesting rewards from the \sidechain. In order to reproduce such environment, we deployed several AWS EC2 \texttt{t2.large} instances (2~vCPUs, 8~GB~RAM). We performed measurements by running 10, 30, 60, and 100 concurrent clients which request rewards from the \sidechain at roughly the same time. Using this setup, we measure the time it takes for individual clients to complete the reward calculation. In addition, we measure the \mbox{end-to-end} performance of the protocol.

%% file: Figs/fig-several-users.tex
\begin{figure}[t]
    \centering
\begin{tikzpicture}[scale=0.8,
  every axis/.style={ 
    ybar stacked,
    ymin=0,ymax=10,
    ylabel=Time (seconds),
    symbolic x coords={
      10 users, ,
      30 users, ,
      60 users, ,
      100 users, 
    },
  bar width=7pt
  },
]

\begin{axis}[bar shift=-10pt,name=boundary]
\addplot+[fill=yellow!50!gray] coordinates
{(10 users,1.53) (30 users,1.54) (60 users,1.54) (100 users, 1.553)};\label{64}
\addplot+[fill=red!50!gray] coordinates
{(10 users,0.34) (30 users,0.33) (60 users,0.33) (100 users, 0.34)};\label{overhead}
\end{axis}

\begin{axis}[hide axis]
\addplot+[fill=blue!50!gray] coordinates
{(10 users,2.59) (30 users,2.59) (60 users,2.59) (100 users, 2.59)};\label{128}
\addplot+[fill=red!50!gray] coordinates
{(10 users,0.355) (30 users,0.36) (60 users,0.35) (100 users, 0.34)};
\end{axis}

\begin{axis}[bar shift=10pt,hide axis]
\addplot+[fill=green!50!gray] coordinates
{(10 users,4.69) (30 users,4.7) (60 users,4.69) (100 users, 4.69)};\label{256}
\addplot+[fill=red!30!gray] coordinates
{(10 users,0.34) (30 users,0.355) (60 users,0.35) (100 users, 0.36)};
\end{axis}

\node[draw, fill=white,inner sep=2pt,above=-0.97cm, left=-2.85cm, scale=0.7] at (boundary.north west) {$
\begin{array}{cl}
    \multicolumn{2}{c}{\text{ Reward request for:}}\\
    \ref{64} & 64 \text{ ads}\\
    \ref{128} & 128 \text{ ads}\\
    \ref{256} & 256 \text{ ads}\\
    \hline
    \ref{overhead} & \text{ Proof gen. and verif.} \\ 
    \end{array}$};
\end{tikzpicture}\vspace{-0.2cm}

    \caption{Cumulative time for \THEMIS protocol to run with~64 (yellow),~128 (blue), and~256 (green) ads. In red, the overhead caused by proving and verifying correct decryption. Results presented for number of concurrent clients requesting the reward computation from the smart contract.}
    \label{fig:themis-several-users}
\end{figure}
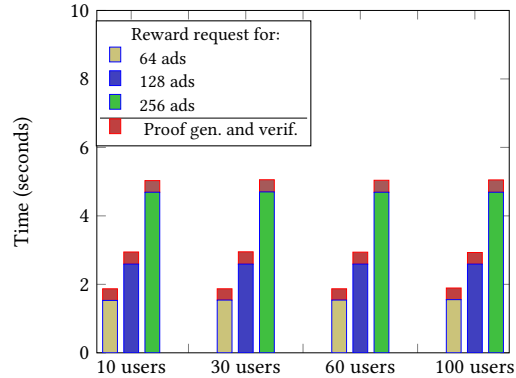

%% file: Figs/fig-multi-sidechains.tex
\begin{figure}[t]
    \centering
\begin{tikzpicture}[scale=0.75,
  every axis/.style={ 
    ymin=0,ymax=320,
    ylabel=Millions Users per Month,
    xlabel=Number of Sidechains,
  bar width=20pt
  },
]

\begin{axis}[
name=boundary, 
legend pos=north west
]
\addplot[color=blue!50!gray, mark=x,thick] coordinates
{(1,51) (2,100) (3,153)};
\addlegendentry{Experimental results};
\addplot[dotted,color=red, mark=x,thick] coordinates
{(3, 153) (4,200) (5,251) (6,306)};
\addlegendentry{Estimated results};
\end{axis}

\end{tikzpicture}
    \caption{Number of Millions of Users that \THEMIS can handle per month by deploying \sidechains. The expected linear growth is confirmed by our experiments.}
    \label{fig:multiple-side-chains}
\end{figure}
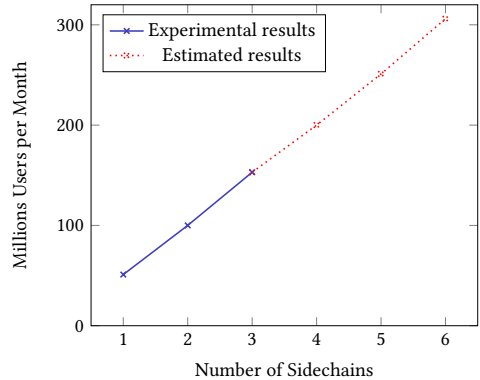

%% file: Sections/07_discussion.tex
\section{Discussion}
\label{discussion}

\subsection{Misbehaving \cf}
\label{sec:misbehavingCF}

In \THEMIS a \cf can cheat in two ways: 
(1)~as it is the entity orchestrating the confidential payments, it may send incorrect rewards to users or, 
(2)~it could use its power to send rewards not only to the user but to other accounts of their control. Both of these actions may be discovered by either users or advertisers.
\begin{enumerate}[leftmargin=0.5cm, label=\Alph*.]
    \item In case of scenario~(1), users can provably challenge \cf for incorrect behaviour by proving that the payment received does not correspond to the payment request they generated. To do so, the user calls the \FSC to prove that the amounts received by the private payment does not correspond to the decrypted aggregate in the payment request $\payreq$. We stress that in case a user must undergo such a scenario, only the aggregate amount of a single \adcatalog will be disclosed (and not its interaction with ads). 
    \item In case of scenario~(2), the escrow account will not have enough funds, resulting in some advertiser getting a smaller refund to what is stated in the performance report of their ad campaign (as described in Section~\ref{sec:aggrcount}). 
    In this case, the advertiser can prove that the received refund does not correspond to the amount staked in Phase~\ref{phase:def-ads-themis} of \THEMIS, minus the rewards paid to users based on the numbers of clicks their ads received.
\end{enumerate}
To claim misbehaviour, users and advertisers can file a complaint via a public function on \FSC (that validates the complaint). If any  complaints are filed, the \FSC switches its state to ``failed'' and \cf will not receive any processing fees, something that affects their reputation.

\subsection{Client-side Fraud}
\label{clientside_fraud}
In \THEMIS, the user agent is responsible to keep the state of ad interactions over time and, locally, to assemble a reward request based on the local state (Section~\ref{sec:demis}). Although this design decentralizes the protocol and distributes trust across all actors of the protocol, it opens possibility for ill intent actors to erroneously claim ad interactions in order to optimize their financial gains. In the extreme, an adversary could assemble a synthetic ad interaction vector and proceed to claim the reward (Section~\ref{phase:claimrewards-themis}), even without having interacted with any ad. Another \mbox{client-side} attack vector in \THEMIS consists of adversaries running multiple Sybil nodes to increase their ad rewards.

Client-side fraud is a hard problem that has been thoroughly researched. There are several practical mechanisms to address client-side fraud (\ie bot clicks, click farms, sybil attacks), that can be used in the context of \THEMIS  (\eg distributed user reputation systems~\cite{yang2019decentralized}, anomaly detection, bluff ads~\cite{haddadi2010fighting}, bio-metric systems~\cite{2019zksense}, client puzzles~\cite{4215910}, \etc) that can be used in the context of \THEMIS. In this section, we outline strategies that reduce the incentives for clients to cheat and to control the number of sybils that an adversary is able to run in the system.

\point{Protections Against Sybil Attacks} Most financial application are required by law to request users to undergo Know Your Customer (KYC) procedures, where users are required to confirm their humanity by providing legal identity proof of humanity. This process makes it harder~--~if not infeasible~--~to run multiple nodes that have successfully undergo KYC. In the context of \THEMIS, the \cf may enforce that only users that have been KYC'ed can successfully request rewards, which can be enforced at the smart contract level using \mbox{privacy-preserving} KYC protocols~\cite{kycethereum}. \mbox{On-chain} \mbox{privacy-preserving} KYC systems work by enforcing users to prove a valid KYC thorugh \mbox{zero-knowledge} proofs that are verified on \mbox{on-chain}. This proof can be integrated in the \FSC~\ref{sec:demis} before proceeding to the reward payment (Section~\ref{sec:demis}).
In addition to \mbox{KYC-based} protections, Sybil attacks can be further attenuated by distributed user reputation systems~\cite{yang2019decentralized}, bio-metric systems~\cite{2019zksense} and client puzzles~\cite{4215910} mechanisms.

\point{Protection Against Synthetic Ad Interactions}
The synthetic ad interactions attack consist of users assembling an ad reward request without having interacted with ads. This attack can be prevented by removing the incentives for users to perform it. The intuition behind the protection is that if the attacker is not able to profit enough from creating synthetic ad interactions, she will not perform it.
The protection consists of ensuring that the \FSC only accepts reward requests below a certain threshold, thus limiting how much rewards can be earned through assembling fraudulent reward requests. We believe that this mechanism together with the protection against Sybils will render the synthetic ad interaction attack unappealing to adversaries. However, we leave improving the protection against synthetic attacks as a research topic for the future.  

%% file: Sections/08_related.tex
\section{Related Work}
\label{sec:related}
The current advertising ecosystem abounds with issues associated with its performance, its transparency, the user's privacy and the integrity of billing and reporting. These failures are already well studied and there are numerous works aiming to shed light on how digital advertising works~\cite{goldstein2013cost,vallina2012breaking,pachilakis2019no,gill2013best,reznichenko2011auctions,rtbPrices17}.

Apart from the studies highlighting the failures of current ad delivery protocols 
there are also important novel ad systems proposed. 
In~\cite{10.5555/646139.680791}, Juels is the first to study private targeted 
advertising. Author proposes a privacy-preserving targeted ad delivery scheme 
based on PIR and Mixnets. In this scheme, advertisers choose a negotiant function that assigns the most fitting ads in their database for each type of profile.  The proposed scheme relies on heavy cryptographic operations and therefore it suffers from intensive computation cost. Their approach focuses on the private distribution of ads and does not take into account other aspects such as view/click reporting. 

In~\cite{toubiana2010adnostic}, authors propose Adnostic: an architecture to enable users to retrieve ads on the fly. Adnostic prefetches \emph{n} ads before the user starts browsing and stores them locally. Aside from the performance benefits of this strategy, Adnostic does this prefetching also in order to preserve the privacy of the user. The parameter \emph{n} is configurable: larger \emph{n} means better ad matching, when smaller \emph{n} means less overhead. In order for the ad-network to correctly charge the corresponding advertisers, Adnostic performs secure billing by using homomorphic encryption and zero-knowledge proofs.

In~\cite{haddadi2009not,privad,reznichenko2014private}, authors propose Privad: an online ad system that aims to be faster and more private than today's ad schema. Privad introduces an additional entity called Dealer. The Dealer is responsible for anonymizing the client so as to prevent the ad-network from identifying the client and also handle the billing. To prevent the Dealer from accessing user's behavioral profile and activity it encrypts the communications between the client and the Dealer. A limitation of Privad is that Dealer is a centralized entity that needs to be always online. 

In~\cite{backes2012obliviad}, authors propose ObliviAd: a provably secure and
practical online behavioral advertising architecture that relies on a secure remote co-processor (SC) and Oblivious RAM (ORAM) to provide the so called secure hardware-based PIR. In ObliviAd, to fetch an ad, a user first sends their encrypted  behavioral profile to the SC which securely selects the ads that match best based on the algorithm specified by the ad network. To prevent the ad-network from learning which ads are selected, they leverage an ORAM scheme. The selected ads are finally sent to the user encrypted, along with
fresh tokens used to billing. User will send back one of these tokens as soon as they view/click on an ad.

In~\cite{7164916} authors point out that, in current advertising systems
the ad-network exclusively determines the payment to get from advertisers and the revenue to share with publishers. This means that (i) a malicious ad-network can overcharge advertisers or underpay publishers.  To make matters worse, as bills cannot be justified by the ad-network, malicious advertisers
can deny actual views/clicks to ask for refund.  On the other hand, (ii) malicious publishers may claim clicks that did not happen, in order to demand higher revenues. To address this problem of unfairness, authors propose a protocol where the ad click reports are encrypted by the user using the public key of the ad-network and signed by both publishers and advertisers.

In~\cite{green2016protocol}, authors use an additively encryption scheme to design a protocol that enables privacy-preserving advertising reporting at scale, without needing any trusted hardware. Performance evaluation results show that their protocol reduces the overhead of reporting by orders of magnitude compared to the ElGamal-based solution of Adnostic~\cite{toubiana2010adnostic} (\ie  1 MB  of  bandwidth  per impression when handing 32,000 advertisements). Contrary to our approach, authors assume a Trusted Third Party (TTP) that owns the key for the homomorphic encryption.

In~\cite{10.1145/2462456.2464436}, authors propose
CAMEO: a framework for mobile advertising that employs intelligent and proactive prefetching of advertisements. CAMEO uses context prediction, to significantly reduce the bandwidth and energy overheads, and provides a negotiation protocol that empowers applications to subsidize their data traffic costs by ``bartering'' their advertisement rights for access bandwidth from mobile ISPs. 
In~\cite{10.1145/1859983.1859993}, authors propose a location-aware, personalised and private advertising system for mobile platforms. In this system, ads are locally broadcast to users within mobile cells. The ad matching happens locally based on the user interests. Finally ad view and click reports are collected using a DTN system. In~\cite{biswas2014privacy}, authors propose a new ad protocol that uses homomorphic and searchable encryption  to allow users transmit mobile sensor data to a cloud service that responds back with the best matching contextual advertisements. 

In~\cite{10.1145/2534169.2486038}, authors present VEX, a protocol for ad exchanges to run low-latency and high-frequency ad auctions that are verifiable and auditable, in order to prevent fraud in a context where parties participating in the auction -- bidders and ad exchanges -- may not know each other. Based on their evaluation of the system, the authors claim that the additional storage required and latency imposed by VEX are low and practical in the context of ad auctions. 
In~\cite{pang2015prota}, authors present and implement PROTA, a privacy-preserving protocol for \mbox{real-time} advertising which uses keywords to match users interests with ads. By using bloom filters, the authors make the ad matching task efficient. The protocol relies on a trusted third party to cooperate with the ad exchange during the bidding and ad delivering phase. The authors implement and evaluate the protocol, and conclude that the time upper bond for matching ads is 200ms, which is considerable practical in the context of an ad matching system.

In~\cite{5958026}, authors present and evaluate a system that aims at providing high-quality ad targeting in multiple scenarios, while giving the user the ability to control their privacy. The system consists of tailored extensions that \emph{mine} the user behaviour locally with low overhead. The extensions generate user behavioural data that can be shared with advertisers without leaking undesirable user information. Similarly to \THEMIS, the authors discuss how the system can be used by users and advertisers, and how it can be used as a replacement for the tracking-based business model in the online advertising industry.

In~\cite{8228673}, authors set out to formalize the concept of privacy in the context of the online advertising ecosystem and to develop a provably secure privacy-preserving protocol for the online advertising ecosystem. While the authors claim that the definition of privacy presented in the paper is more useful compared to previous work in the online advertising context, their attempts to develop a provably secure privacy-preserving protocol has failed due to being hard to balance privacy with usefulness of the user data. The authors conjecture that cryptographic mechanisms have the potential to solve the privacy versus data usefulness conundrum. Using applying cryptography is the basis of how \THEMIS proposes to preserve privacy when calculating ad rewards, providing advertisers with campaign metrics and performing confidential payments to users.

Towards a similar direction with the user rewarding schema of \THEMIS, in~\cite{wang2015privacy}, authors
propose a privacy-aware framework to promote targeted advertising. In this framework, an ad broker responsible for handling ad targeting, sits between advertisers and users and provides certain amount of compensation to incentivize users to click ads that are interesting yet sensitive to them. 
In~\cite{parra2017pay}, authors propose a targeted advertising framework which enables users to get compensated based on the amount of user tracking they sustain and the privacy they lose. The authors analyze the interaction between the different parties in the online advertising context --- advertisers, the ad broker and users --- and propose a framework where the interactions between the different parties are a positive-sum game. In this game, all parties are incentivized to behave according to what other parties expect, achieving an equilibrium where everyone benefits. More specifically, the users determine their click behaviour based on their interested and their privacy leakage, which in turn will influence the advertisers and ad broker to provide less invasive and better ads. \THEMIS relies on a similar game theoretical approach. By providing compensation for good behaviour while providing the verification mechanisms for all parties to audit whether everyone is behaving according to the protocol, the incentives to cheat and misbehave are lower.

%% file: Sections/09_conclusions.tex
\section{Conclusions}
\label{sec:conclusions}
In this paper, we presented \THEMIS, the first decentralized, scalable,  and privacy-preserving ad platform that provides integrity and auditability to its participants, so users do not need to blindly trust any of the protocol actors. To increase the user engagement with ads and provide advertisers with the necessary performance feedback about their ad campaigns, \THEMIS (i) rewards users for their ad viewing and (ii) provides advertisers with verifiable and privacy-preserving campaign performance analytics.

We implemented our approach by leveraging a permissioned blockchain with Solidity smart contracts as well as zero-knowledge techniques. We evaluate the scalability and performance of our prototype and show that \THEMIS can support reward payments for more than~\usersScalability users per month on a single-\sidechain setup and~\usersScalabilityMulti  users on a parallel multi-\sidechain setup, proving linear scalability.

While many projects and companies have proposed the use of blockchain for online advertising, we believe that \THEMIS is the first system that delivers on that promise. Given the practicality of the approach and the combination of security, privacy, and performance properties it delivers, \THEMIS can be used as a foundation of a radically new approach to online advertising. 




%% file: Sections/10_appendix_smart_contract.tex
\section{Smart Contracts}
\label{sec:appendix-smart-contracts}

\newcommand{\scfunction}[1]{\underline{\texttt{#1}}}
\newcommand{\nrpolicies}{l}

\newcommand{\fundslist}{\mathcal{F}}
\newcommand{\fundi}[1]{\fundslist_{#1}}
\newcommand{\tokensamount}{\tau}
\newcommand{\paymentrequests}{\texttt{PaymentReq}}
\newcommand{\payedrequests}{\texttt{PayedReq}}
\newcommand{\processingfees}{\omega}
\newcommand{\campaigninit}{\texttt{Init}}
\newcommand{\txref}{\texttt{TxRef}}

In this section, we specify the functionality and properties of the smart contracts necessary to run \THEMIS. In practice and at the EVM level, the policy smart contract logic and fund smart contract logic may be split into multiple smart contracts, but for the sake of simplicity, we describe the logic as part of two smart contracts per \THEMIS campaign: (i) the Policy Smart Contract (Figure~\ref{fig:psm}), and (ii) the Fund Smart Contract (Figure~\ref{fig:fsm}). We assume that both smart contracts have access to storage of the $\cf$'s public key and the consensus participants public key (when generated).

\subsection{Policy Smart Contract}
\point{Public data structures} The Policy Smart Contract (\PSC) keeps its state in three public data structures: an array with the encrypted ad policies ($\encpolicy$), an array containing all rewards aggregates (\texttt{Agg}) calculated by the smart contract, and an array of encrypted symmetric keys, $\encseckey$, used to encrypt the policies. The latter allows the validators to decrypt the policies and apply them to the aggregate requests. 

\point{$\texttt{StorePolicy()}$} This private function can be called only by the account which deployed the smart contract, \ie by the Campaign Facilitator (\cf). The function receives an array of $\texttt{uint265}$ types -- which represent the encrypted ad policies for the campaign agreed with the advertisers (Phase~\ref{phase:def-ads-themis} in Section~\ref{sec:demis}) --- and it initializes the public ($\encpolicy$) data structure with its input. Each policy is encrypted using a symmetric key agreed between the \cf and the corresponding advertiser.

\point{$\texttt{ComputeAggregate()}$} It is a public function that is exposed to the users. Users call this function with an array containing their encrypted interactions (Phase~\ref{phase:claimrewards-themis} in Section~\ref{sec:demis}). The smart contract proceeds to calculate the reward aggregate based on the user input and the ads policies ($\encpolicy$). The aggregate is stored in the (\texttt{Agg}) data structure, which is accessible to all users.

\point{$\texttt{GetAggregate()}$} It is a public function which receives an ID (\ie $\texttt{uint265}$) and returns the encrypted aggregate indexed by the respective ID in the \texttt{Agg} data structure. This function is used by users to fetch the encrypted aggregate calculated by the smart contract, after having called the $\texttt{ComputeAggregate()}$ function described above.

\point{$\texttt{PaymentRequest()}$} It is a public function which receives an encrypted payment request from users (Phase~\ref{phase:paymentrequest-themis} in Section~\ref{sec:demis}) under the validators' key. If the payment request is valid, the smart contract buffers the request in the Fund Smart Contract ($\paymentrequests$). The buffered payments are settled periodically by \cf.

\input{Sections/appendix_code/pcm}

\subsection{Fund Smart Contract}
\point{Public data structures} The Fund Smart Contract (\FSC) keeps it state in multiple public data structures. The $\campaigninit$ parameter represents whether the ad campaign has started. In order for the $\campaigninit$ to turn to be marked as initialized (\ie $true$), all campaign advertisers -- which are kept in $\advs$ -- must confirm their participation by depositing the funds (to the ad campaign's escrow account $\fundslist$) necessary to cover their ad campaign in the smart contract account. In addition, the \FSC keeps a list of all payment requests triggered by \PSC and the successfully payed requests $\payedrequests$. Finally it stores the agreed processing fees of the \cf, and a value of the overall ad interaction, $\aggrclicks$, which is updated by the consensus pool.

\point{$\texttt{StoreAdvID()}$} Private function that can be called by the \cf to add new advertisers to the campaign. If the campaign has not started (\ie $\campaigninit: false$), the advertisers ID is added to the $\advs$ list. No new advertisers can be added after a campaign has started.

\point{$\texttt{StoreAggrClicks()}$} This public function has an access control policy that only a value signed by the consensus participants will update the public data structure. It is a function that is used to update the state of the smart contract overall ad interactions, $\aggrclicks$.

\point{$\texttt{StoreFunds()}$} Public function called by the advertisers upon transferring the campaign funds to the \FSC account. When all the advertisers have transferred the funds necessary to cover their ad campaign, the smart contract updates its state to $\campaigninit: true$.

\point{$\texttt{InitialiseCampaign()}$} Private function that is only called by the smart contract. It sets $\campaigninit: true$.

\point{$\texttt{SettlementRequest()}$} This public function has an access control policy that only the \cf can successfully call it. It requests the release of funds to the \cf account, in order for the \cf to be able to have the funds for settling the buffered payment requests (in \PSC).

\input{Sections/appendix_code/fcm}

\point{$\texttt{RefundAdvertisers()}$} This function is called internally by the \FSC and it is triggered either when: (1) all the ads in the campaign have been "spent" by the users or (2) when a pre-defined \textit{epoch} has passed, signaling the end of the campaign. This function releases the funds to the advertisers, based on the $\aggrclicks$ per advertiser in the campaign.

\point{$\texttt{PayProcessingFees()}$} This function is -- alongside with $\texttt{RefundAdvertisers()}$ -- called when the campaign is finished. It verifies if the \cf has behaved correctly and pays the fees to the \cf's account.

\point{$\texttt{RaiseComplaint()}$} This public function allows users to prove that the \cf misbehaved. In order to cull such a function, users must prove that their corresponding aggregate does not correspond to the private payment they received. To this end they disclose the aggregate value they were expecting to earn. This will prove that the \cf misbehaved and the smart contract can flag the latter as such.

\point{$\texttt{ClaimInsufficientRefund()}$} This public function allows advertisers to prove they have received insufficient refunds. To this end, advertisers simply call this function which automatically checks the validity of the claim. If the claim is valid, it flags the \cf as misbehaving.

%% file: Sections/appendix_code/pcm.tex
\begin{figure}[t]
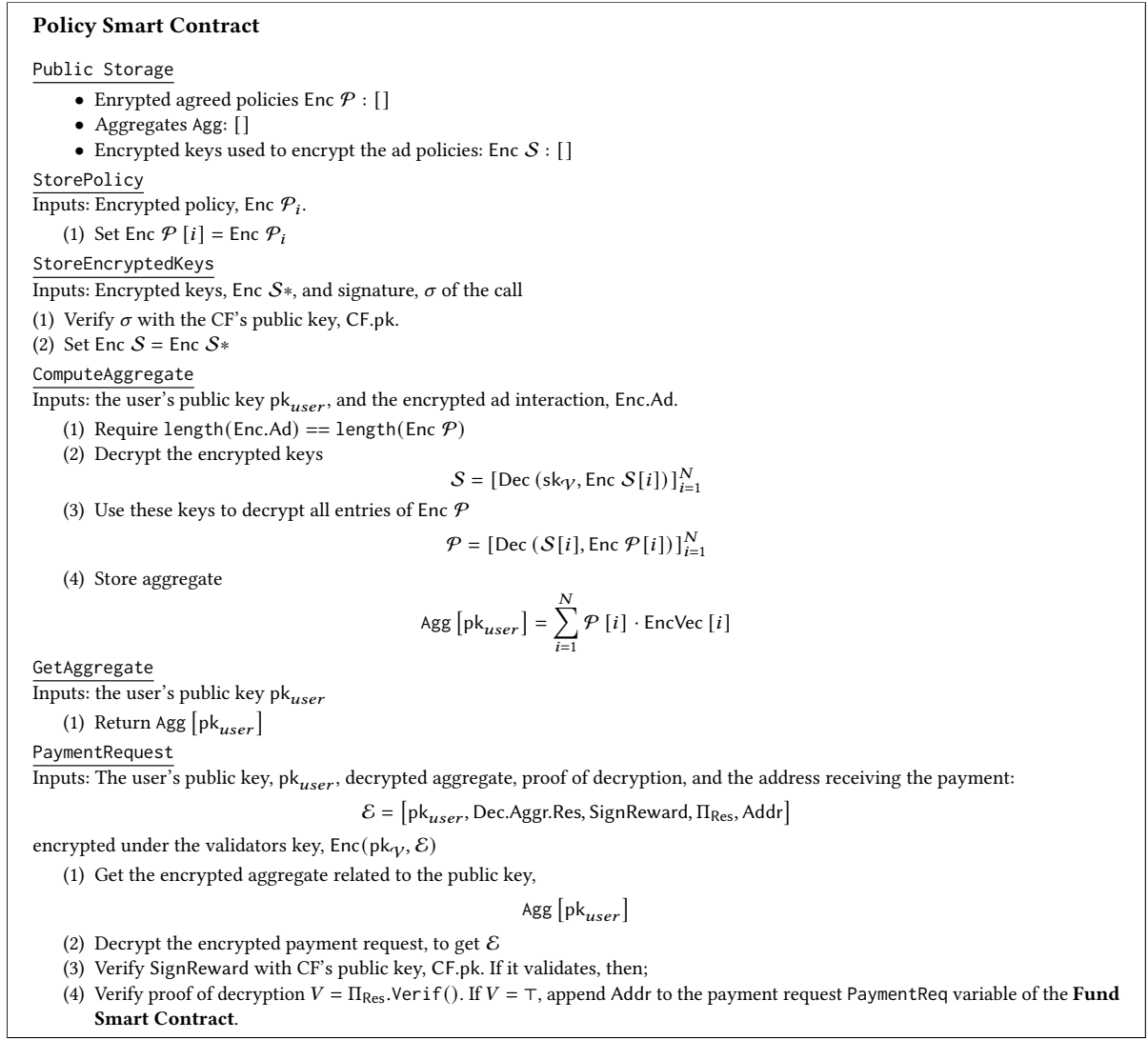

\footnotesize
\begin{cvbox}[Policy Smart Contract]
\scfunction{Public Storage}
\begin{itemize}
    \item Enrypted agreed policies $\encpolicy: \left[\right]$
    \item Aggregates \texttt{Agg}: $\left[\right]$
    \item Encrypted keys used to encrypt the ad policies: $\encseckey:\left[\right]$
\end{itemize}
\scfunction{StorePolicy}

Inputs: Encrypted policy, $\encpolicy_i$.
\begin{enumerate}
    \item Set $\encpolicy\left[i\right] = \encpolicy_i$
\end{enumerate}
\scfunction{StoreEncryptedKeys}

\noindent Inputs: Encrypted keys, $\encseckey*$, and signature, $\signature$ of the call
\begin{enumerate}[leftmargin=*]
    \item Verify $\signature$ with the \cf's public key, $\cfpk$.
    \item Set $\encseckey=\encseckey*$
\end{enumerate}

\underline{\texttt{ComputeAggregate}}

\noindent Inputs: the user's public key $\pk_{user}$, and the 
encrypted ad interaction, $\encryptedad$.
\begin{enumerate}
    \item Require 
    $\texttt{length}(\encryptedad) == \texttt{length}(\encpolicy)$
    \item Decrypt the encrypted keys 
    \[\seckey = \left[\dec\left(\sk_{\mathcal{V}},\encseckey[i]\right)\right]_{i=1}^N\] 
    \item Use these keys to decrypt all entries of $\encpolicy$
    \[ \policy = \left[\dec\left(\seckey[i],\encpolicy[i]\right)\right]_{i=1}^N\]
    \item Store aggregate
    \[\texttt{Agg}\left[\pk_{user}\right] = \sum_{i=1}^{N} \policy\left[i\right] \cdot \encvec\left[i\right]\]
\end{enumerate}
\scfunction{GetAggregate}

\noindent Inputs: the user's public key $\pk_{user}$
\begin{enumerate}
    \item Return $\texttt{Agg}\left[\pk_{user}\right]$
\end{enumerate}
\scfunction{PaymentRequest}

\noindent Inputs: The user's public key, $\pk_{user}$, decrypted aggregate, proof of decryption, and the address receiving the payment:
\[
\payreq=\left[\pk_{user},\decryptedaggr, \signaturereward, \aggrproofdec, \addr \right]
\]
encrypted under the validators key, $\enc(\valikey, \payreq)$
\begin{enumerate}
    \item Get the encrypted aggregate related to the public key, 
    \[\texttt{Agg}\left[\pk_{user}\right]\]
    \item Decrypt the encrypted payment request, to get $\payreq$
    \item Verify $\signaturereward$ with \cf's public key, $\cfpk$. If it validates, then;
    \item Verify proof of decryption $V=\aggrproofdec.\texttt{Verif}()$. If $V=\top$, append $\addr$ to the payment request $\paymentrequests$ variable of the \textbf{Fund Smart Contract}.
\end{enumerate}
\end{cvbox}
\caption{Description of the public storage and functionality of the Policy Smart Contract (PSC)}
\label{fig:psm}
\end{figure}

%% file: Sections/appendix_code/fcm.tex
\begin{figure*}[t]
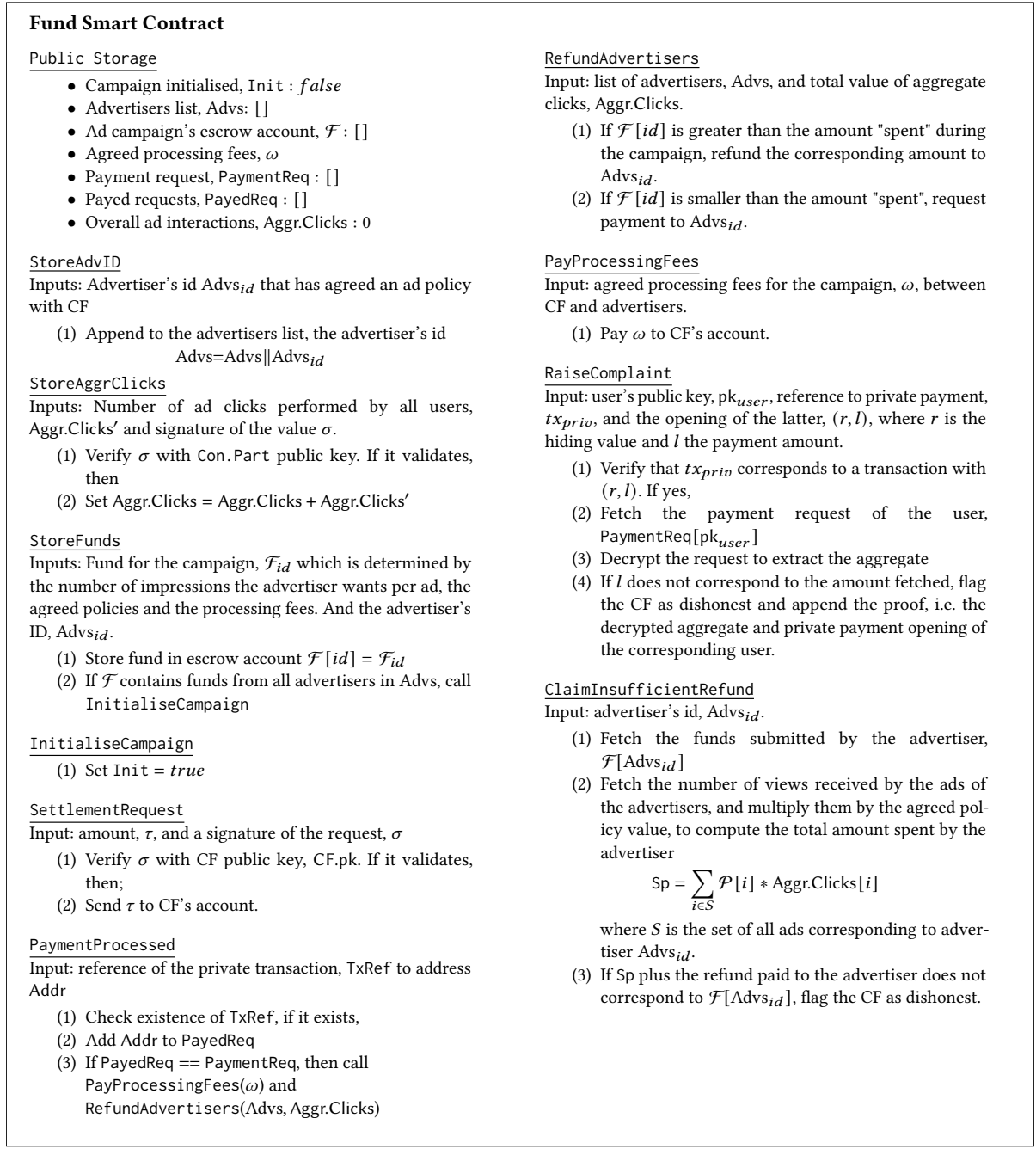

\begin{cvbox}[Fund Smart Contract]
\begin{minipage}[t]{.45\textwidth}
\scfunction{Public Storage}
\begin{itemize}
    \item Campaign initialised, $\campaigninit: false$
    \item Advertisers list, $\advs: \left[\right]$
    \item Ad campaign's escrow account, $\fundslist: \left[\right]$
    \item Agreed processing fees, $\processingfees$
    \item Payment request, $\paymentrequests: \left[\right]$
    \item Payed requests, $\payedrequests: \left[\right]$
    \item Overall ad interactions, $\aggrclicks: 0$
\end{itemize}
\vspace{2mm}
\scfunction{StoreAdvID}

\noindent Inputs: Advertiser's id $\advs_{id}$ that has agreed an ad policy with \cf
\begin{enumerate}
    \item Append to the advertisers list, the advertiser's id 
    
    \centerline{$\advs = \advs \| \advs_{id}$}
\end{enumerate}
\scfunction{StoreAggrClicks}

\noindent Inputs: Number of ad clicks performed by all users, $\aggrclicks'$ and signature of the value $\signature$. 
\begin{enumerate}
    \item Verify $\signature$ with $\cp$ public key. If it validates, then
    \item Set $\aggrclicks = \aggrclicks + \aggrclicks'$
\end{enumerate}
\vspace{2mm}
\scfunction{StoreFunds}

\noindent Inputs: Fund for the campaign, $\fundi{id}$ which is determined by the number of impressions the advertiser wants per ad, the agreed policies and the processing fees. And the advertiser's ID, $\advs_{id}$.
\begin{enumerate}
    \item Store fund in escrow account $\fundslist\left[id\right] = \fundi{id}$
    \item If $\fundslist$ contains funds from all advertisers in $\advs$, call 
    \item[] \texttt{InitialiseCampaign}
\end{enumerate}
\vspace{2mm}
\scfunction{InitialiseCampaign}

\begin{enumerate}
    \item Set $\campaigninit = true$
\end{enumerate}
\vspace{2mm}
\scfunction{SettlementRequest}

\noindent Input: amount, $\tokensamount$, and a signature of the request, $\sigma$
\begin{enumerate}
    \item Verify $\sigma$ with \cf public key, $\cfpk$. If it validates, then;
    \item Send $\tokensamount$ to \cf's account.  
\end{enumerate}
\vspace{2mm}
\scfunction{PaymentProcessed}

\noindent Input: reference of the private transaction, $\txref$ to address $\addr$
\begin{enumerate}
    \item Check existence of $\txref$, if it exists,
    \item Add $\addr$ to $\payedrequests$
    \item If $\payedrequests == \paymentrequests$, then call 
    \item[] \texttt{PayProcessingFees}($\processingfees$) and
    \item[] \texttt{RefundAdvertisers}($\advs, \aggrclicks$)
\end{enumerate}
\vspace{2mm}
\end{minipage}
\hspace{1cm}
\begin{minipage}[t]{.45\textwidth}
\scfunction{RefundAdvertisers}

\noindent Input: list of advertisers, $\advs$, and total value of aggregate clicks, $\aggrclicks$.
\begin{enumerate}
    \item If $\fundslist\left[id\right]$ is greater than the amount "spent" during the campaign, refund the corresponding amount to $\advs_{id}$.
    \item If  $\fundslist\left[id\right]$ is smaller than the amount "spent", request payment to $\advs_{id}$.
\end{enumerate}
\vspace{2mm}
\scfunction{PayProcessingFees}

\noindent Input: agreed processing fees for the campaign, $\processingfees$, between \cf and advertisers.
\begin{enumerate}
    \item Pay $\processingfees$ to \cf's account.
\end{enumerate}
\vspace{2mm}
\scfunction{RaiseComplaint}

\noindent Input: user's public key, $\pk_{user}$, reference to private payment, $tx_{priv}$, and the opening of the latter, $(r, l)$, where $r$ is the hiding value and $l$ the payment amount.
\begin{enumerate}
\item Verify that $tx_{priv}$ corresponds to a transaction with $(r, l)$. If yes, 
\item Fetch the payment request of the user, $\paymentrequests[\pk_{user}]$
\item Decrypt the request to extract the aggregate
\item If $l$ does not correspond to the amount fetched, flag the \cf as dishonest and append the proof, i.e. the decrypted aggregate and private payment opening of the corresponding user.
\end{enumerate}  
\vspace{2mm}
\scfunction{ClaimInsufficientRefund}

\noindent Input: advertiser's id, $\advs_{id}$.
\begin{enumerate}
\item Fetch the funds submitted by the advertiser, $\fundslist[\advs_{id}]$
\item Fetch the number of views received by the ads of the advertisers, and multiply them by the agreed policy value, to compute the total amount spent by the advertiser
\[\texttt{Sp} = \sum_{i\in S}\policy[i] * \aggrclicks[i]\]
where $S$ is the set of all ads corresponding to advertiser $\advs_{id}$. 
\item If $\texttt{Sp}$ plus the refund paid to the advertiser does not correspond to $\fundslist[\advs_{id}]$, flag the \cf as dishonest.
\end{enumerate}
\end{minipage}
\end{cvbox}
\caption{Description of the public storage and functionality of the Fund Smart Contract (FSM)}
\label{fig:fsm} 
\end{figure*}

%% file: main.bbl
\begin{thebibliography}{10}

\bibitem{paywalls}
Panagiotis Papadopoulos, Peter Snyder, and Benjamin Livshits.
\newblock Keeping out the masses: Understanding the popularity and implications
  of internet paywalls.
\newblock In {\em Proceedings of the The World Wide Web Conference}, WWW'20.
  ACM, 2020.

\bibitem{wikipediaFunds1}
Jeff Dunn.
\newblock Wikipedia is asking for donations again — here’s how much cash it
  already has in the bank.
\newblock
  \url{https://www.businessinsider.com/wikipedia-donations-profit-money-chart-2016-11},
  2016.

\bibitem{truth2018}
Panagiotis Papadopoulos, Panagiotis Ilia, and Evangelos Markatos.
\newblock Truth in web mining: Measuring the profitability and the imposed
  overheads of cryptojacking.
\newblock In {\em Information Security}, 2019.

\bibitem{10.1145/1963405.1963451}
Ghassan~O. Karame, Aur\'{e}lien Francillon, and Srdjan \v{C}apkun.
\newblock Pay as you browse: Microcomputations as micropayments in web-based
  services.
\newblock In {\em Proceedings of the 20th International Conference on World
  Wide Web}, WWW ’11, 2011.

\bibitem{frag1}
Muhammad~Ahmad Bashir, Sajjad Arshad, William Robertson, and Christo Wilson.
\newblock Tracing information flows between ad exchanges using retargeted ads.
\newblock In {\em Proceedings of the 25th USENIX Conference on Security
  Symposium}, 2016.

\bibitem{frag2}
Panagiotis Papadopoulos, Nicolas Kourtellis, and Evangelos~P. Markatos.
\newblock Cookie synchronization: Everything you always wanted to know but were
  afraid to ask.
\newblock In {\em Proceedings of The World Wide Web Conference}, 2019.

\bibitem{fraud1}
N.~{Kshetri}.
\newblock The economics of click fraud.
\newblock {\em IEEE Security Privacy}, 2010.

\bibitem{fraud2}
S.~{Kumari}, X.~{Yuan}, J.~{Patterson}, and H.~{Yu}.
\newblock Demystifying ad fraud.
\newblock In {\em 2017 IEEE Frontiers in Education Conference (FIE)}, 2017.

\bibitem{adfraud}
Michael Burgi.
\newblock What's being done to rein in \$7 billion in ad fraud.
\newblock
  \url{https://www.adweek.com/brand-marketing/whats-being-done-rein-7-billion-ad-fraud-169743/},
  2016.

\bibitem{Zarras:2014:DAM:2663716.2663719}
Apostolis Zarras, Alexandros Kapravelos, Gianluca Stringhini, Thorsten Holz,
  Christopher Kruegel, and Giovanni Vigna.
\newblock The dark alleys of madison avenue: Understanding malicious
  advertisements.
\newblock In {\em Proceedings of the 2014 Conference on Internet Measurement
  Conference}, IMC '14, 2014.

\bibitem{malvertising}
Jerome Dangu.
\newblock Uncovering 2017’s largest malvertising operation.
\newblock
  \url{https://blog.confiant.com/uncovering-2017s-largest-malvertising-operation-b84cd38d6b85},
  2018.

\bibitem{middlemenHalf}
Alex Barker.
\newblock Half of online ad spending goes to industry middlemen.
\newblock
  \url{https://www.ft.com/content/9ee0ebd3-346f-45b1-8b92-aa5c597d4389}, 2020.

\bibitem{trackersWWW2016}
Elias~P Papadopoulos, Michalis Diamantaris, Panagiotis Papadopoulos, Thanasis
  Petsas, Sotiris Ioannidis, and Evangelos~P Markatos.
\newblock The long-standing privacy debate: Mobile websites vs mobile apps.
\newblock In {\em Proceedings of the 26th International Conference on World
  Wide Web}, 2017.

\bibitem{exclusiveCSync}
Panagiotis Papadopoulos, Nicolas Kourtellis, and Evangelos~P. Markatos.
\newblock Exclusive: How the (synced) cookie monster breached my encrypted vpn
  session.
\newblock In {\em Proceedings of the 11th European Workshop on Systems
  Security}, EuroSec, 2018.

\bibitem{englehardt2016online}
Steven Englehardt and Arvind Narayanan.
\newblock Online tracking: A 1-million-site measurement and analysis.
\newblock In {\em Proceedings of the 2016 ACM SIGSAC Conference on Computer and
  Communications Security}, 2016.

\bibitem{ctrOverall}
{CXL}.
\newblock What is a “good” click-through rate? click-through rate
  benchmarks.
\newblock \url{https://cxl.com/guides/click-through-rate/benchmarks/}, 2020.

\bibitem{adblockreport2019}
Daniyal Malik.
\newblock Global ad-blocking behaviors in 2019 - stats \& consumer trends.
\newblock
  \url{https://www.digitalinformationworld.com/2019/04/global-ad-blocking-behaviors-infographic.html},
  2019.

\bibitem{adblockersCost}
Dan Shewan.
\newblock The rise of ad blockers: Should advertisers be panicking?(!!).
\newblock \url{https://www.wordstream.com/blog/ws/2015/10/02/ad-blockers},
  2019.

\bibitem{privad}
Saikat Guha, Alexey Reznichenko, Kevin Tang, Hamed Haddadi, and Paul Francis.
\newblock Serving ads from localhost for performance, privacy, and profit.
\newblock In {\em HotNets}, pages 1--6, 2009.

\bibitem{toubiana2010adnostic}
Vincent Toubiana, Arvind Narayanan, Dan Boneh, Helen Nissenbaum, and Solon
  Barocas.
\newblock Adnostic: Privacy preserving targeted advertising.
\newblock In {\em Proceedings Network and Distributed System Symposium}, 2010.

\bibitem{Brave2017}
{Brave Software Inc}.
\newblock {Brave - BAT - Whitepaper}.
\newblock
  \url{https://basicattentiontoken.org/BasicAttentionTokenWhitePaper-4.pdf},
  2017.

\bibitem{ctrBrave}
{Brave Software Inc.}
\newblock Brave reaches 8 million monthly active users and delivers nearly 400
  privacy-preserving ad campaigns.
\newblock
  \url{https://brave.com/brave-reaches-8-million-monthly-active-users-and-delivers-nearly-400-privacy-preserving-ad-campaigns/},
  2019.

\bibitem{7164916}
J.~{Hua}, A.~{Tang}, and S.~{Zhong}.
\newblock Advertiser and publisher-centric privacy aware online behavioral
  advertising.
\newblock In {\em 2015 IEEE 35th International Conference on Distributed
  Computing Systems}, June 2015.

\bibitem{consensus_sok}
Juan~A. Garay and Aggelos Kiayias.
\newblock Sok: {A} consensus taxonomy in the blockchain era.
\newblock In Stanislaw Jarecki, editor, {\em Topics in Cryptology - {CT-RSA}
  2020 - The Cryptographers' Track at the {RSA} Conference 2020, San Francisco,
  CA, USA, February 24-28, 2020, Proceedings}, volume 12006 of {\em Lecture
  Notes in Computer Science}, pages 284--318. Springer, 2020.

\bibitem{EIP650}
{Yu-Te Lin}.
\newblock Istanbul byzantine fault tolerance.
\newblock \url{https://github.com/ethereum/EIPs/issues/650}, 2017.

\bibitem{ibft}
Roberto Saltini.
\newblock Correctness analysis of {IBFT}.
\newblock {\em CoRR}, abs/1901.07160, 2019.

\bibitem{clique}
{IBFT} ({C}lique {C}onsensus {M}echanism).
\newblock
  \url{https://github.com/ethereum/go-ethereum/blob/master/consensus/clique/clique.go},
  2018.

\bibitem{quorumpoa}
Quorum {P}o{A} {N}etwork.
\newblock \url{ https://www.goquorum.com/}.
\newblock Accessed: 07-2019.

\bibitem{quorumgo}
Golang {I}mplementation of {Q}uorum {C}lients.
\newblock \url{ https://github.com/jpmorganchase/quorum}.
\newblock Accessed: 07-2019.

\bibitem{pegasys}
Pegasys {P}o{A}.
\newblock \url{ https://pegasys.tech/}.
\newblock Accessed: 07-2019.

\bibitem{elgamalcrypto}
T.~{Elgamal}.
\newblock A public key cryptosystem and a signature scheme based on discrete
  logarithms.
\newblock {\em IEEE Transactions on Information Theory}, 1985.

\bibitem{Goldwasser:1985:KCI:22145.22178}
S~Goldwasser, S~Micali, and C~Rackoff.
\newblock {The Knowledge Complexity of Interactive Proof-systems}.
\newblock In {\em STOC}, 1985.

\bibitem{Gennaro2007}
Rosario Gennaro, Stanislaw Jarecki, Hugo Krawczyk, and Tal Rabin.
\newblock Secure distributed key generation for discrete-log based
  cryptosystems.
\newblock {\em J. Cryptology}, 2007.

\bibitem{micalivrf}
Silvio Micali, Michael Rabin, and Salil Vadhan.
\newblock Verifiable random functions.
\newblock In {\em 40th Annual Symposium on Foundations of Computer Science},
  1999.

\bibitem{irtf-cfrg-vrf-05}
Sharon Goldberg, Leonid Reyzin, Dimitrios Papadopoulos, and Jan Včelák.
\newblock {Verifiable Random Functions (VRFs)}.
\newblock Internet-Draft draft-irtf-cfrg-vrf-05, Internet Engineering Task
  Force, 2019.

\bibitem{aztecpaper}
The {AZTEC} protocol white paper.
\newblock \url{https://github.com/AztecProtocol/AZTEC/blob/develop/AZTEC.pdf}.
\newblock Accessed: 07-2019.

\bibitem{zether}
Benedikt B{\"{u}}nz, Shashank Agrawal, Mahdi Zamani, and Dan Boneh.
\newblock Zether: Towards privacy in a smart contract world.
\newblock In {\em Financial Cryptography and Data Security - 24th International
  Conference}, 2020.

\bibitem{zcashpaper}
E.~B. {Sasson}, A.~{Chiesa}, C.~{Garman}, M.~{Green}, I.~{Miers}, E.~{Tromer},
  and M.~{Virza}.
\newblock Zerocash: Decentralized anonymous payments from bitcoin.
\newblock In {\em IEEE Symposium on Security and Privacy}, 2014.

\bibitem{zcash}
Zcash website.
\newblock \url{ https://z.cash/}.

\bibitem{anonymouszether}
Anonymous {Z}ether extension paper.
\newblock \url{
  https://github.com/jpmorganchase/anonymous-zether/blob/master/docs/AnonZether.pdf}.

\bibitem{hola}
Hola free vpn - unblock any website.
\newblock \url{https://hola.org/}.

\bibitem{varvello2019vpn0}
Matteo Varvello, I{\~n}igo~Querejeta Azurmendi, Antonio Nappa, Panagiotis
  Papadopoulos, Goncalo Pestana, and Ben Livshits.
\newblock Vpn0: A privacy-preserving decentralized virtual private network.
\newblock {\em arXiv preprint arXiv:1910.00159}, 2019.

\bibitem{radioshakData}
Andrea Peterson.
\newblock Bankrupt radioshack wants to sell off user data. but the bigger risk
  is if a facebook or google goes bust.
\newblock
  https://www.washingtonpost.com/news/the-switch/wp/2015/03/26/bankrupt-radioshack-wants-to-sell-off-user-data-but-the-bigger-risk-is-if-a-facebook-or-google-goes-bust/,
  2015.

\bibitem{brokserSell}
Bernard Marr.
\newblock Where can you buy big data? here are the biggest consumer data
  brokers.
\newblock https://www.forbes.com/sites/bernardmarr/2017/
  09/07/where-can-you-buy-big-data-here-are-the-biggest-consumer-data-brokers/,
  2017.

\bibitem{toysmartData}
Matt Richtel.
\newblock F.t.c. moves to halt sale of database at toysmart.
\newblock
  http://www.nytimes.com/2000/07/11/business/ftc-moves-to-halt-sale-of-database-at-toysmart.html,
  2000.

\bibitem{10.5555/646139.680791}
Ari Juels.
\newblock Targeted advertising ... and privacy too.
\newblock In {\em Proceedings of the Conference on Topics in Cryptology: The
  Cryptographer’s Track at RSA}, 2001.

\bibitem{10.1145/2462456.2464436}
Azeem~J. Khan, Kasthuri Jayarajah, Dongsu Han, Archan Misra, Rajesh Balan, and
  Srinivasan Seshan.
\newblock Cameo: A middleware for mobile advertisement delivery.
\newblock In {\em Proceeding of the 11th Annual International Conference on
  Mobile Systems, Applications, and Services}, MobiSys ’13, 2013.

\bibitem{green2016protocol}
Matthew Green, Watson Ladd, and Ian Miers.
\newblock A protocol for privately reporting ad impressions at scale.
\newblock In {\em Proceedings of the 2016 ACM SIGSAC Conference on Computer and
  Communications Security}, 2016.

\bibitem{yang2019decentralized}
Rupeng Yang, Man~Ho Au, Qiuliang Xu, and Zuoxia Yu.
\newblock Decentralized blacklistable anonymous credentials with reputation.
\newblock {\em Computers \& Security}, 2019.

\bibitem{haddadi2010fighting}
Hamed Haddadi.
\newblock Fighting online click-fraud using bluff ads.
\newblock {\em ACM SIGCOMM Computer Communication Review}, 2010.

\bibitem{2019zksense}
Panagiotis Papadopoulos, Inigo~Querejeta Azurmendi, Jiexin Zhang, Matteo
  Varvello, Antonio Nappa, and Benjamin Livshits.
\newblock {ZKSENSE: a Privacy-Preserving Mechanism for Bot Detection in Mobile
  Devices}, 2019.

\bibitem{4215910}
H.~{Rowaihy}, W.~{Enck}, P.~{McDaniel}, and T.~{La Porta}.
\newblock Limiting sybil attacks in structured p2p networks.
\newblock In {\em IEEE 26th IEEE International Conference on Computer
  Communications}, INFOCOM'07, 2007.

\bibitem{BraveAds}
Emil Protalinski.
\newblock Brave rolls out its own ads that pay users a 70\% cut.
\newblock
  \url{https://venturebeat.com/2019/04/24/brave-rolls-out-its-own-ads-that-pay-users-a-70-cut/},
  2019.

\bibitem{DHKE}
W.~Diffie and M.~Hellman.
\newblock New directions in cryptography.
\newblock {\em IEEE Trans. Inf. Theor.}, 2006.

\bibitem{hpke}
Richard Barnes, Karthikeyan Bhargavan, and Christopher~A. Wood.
\newblock {Hybrid Public Key Encryption}.
\newblock Internet-Draft draft-irtf-cfrg-hpke-04, Internet Engineering Task
  Force, May 2020.
\newblock Work in Progress.

\bibitem{dkg}
John Canny and Stephen Sorkin.
\newblock Practical large-scale distributed key generation.
\newblock In {\em Advances in Cryptology - EUROCRYPT}, 2004.

\bibitem{solidity}
Gavin Wood.
\newblock Solidity documentation.
\newblock \url{https://solidity.readthedocs.io/en/v0.6.2/}, 2014.

\bibitem{eip196}
Christian Reitwiessner.
\newblock Eip 196: Precompiled contracts for addition and scalar multiplication
  on the elliptic curve alt\_bn128.
\newblock \url{https://eips.ethereum.org/EIPS/eip-196}.

\bibitem{web3-rust}
Ethereum json-rpc multi-transport client. rust implementation of web3.js
  library.
\newblock \url{https://github.com/tomusdrw/rust-web3}.

\bibitem{curve25519dalek}
Curve25519-dalek rust implementation.
\newblock \url{https://github.com/dalek-cryptography/curve25519-dalek}.
\newblock Accessed: 07-2019.

\bibitem{elgamal-rust}
A pure-rust implementation of elgamal encryption using the ristretto group over
  curve25519.
\newblock \url{https://crates.io/crates/elgamal_ristretto}.
\newblock Accessed: 07-2020.

\bibitem{mjolnir}
Devops tool to rapidly deploy ethereum proof of authority (poa).
\newblock \url{https://github.com/brave-experiments/Mjolnir}.

\bibitem{quorum}
{Nethereum}.
\newblock Quorum - nethereum documentation.
\newblock
  \url{https://docs.nethereum.com/en/latest/ethereum-and-clients/quorum/},
  2019.

\bibitem{BRAVE}
{Brave Software Inc}.
\newblock Brave launches the first advertising platform built on privacy.
\newblock \url{https://brave.com/brave-ads-launch/}, 2019.

\bibitem{kycethereum}
Alex Biryukov, Dmitry Khovratovich, and Sergei Tikhomirov.
\newblock Privacy-preserving kyc on ethereum.
\newblock In {\em Proceedings of 1st ERCIM Blockchain Workshop 2018}, 2018.

\bibitem{goldstein2013cost}
Daniel~G Goldstein, R~Preston McAfee, and Siddharth Suri.
\newblock The cost of annoying ads.
\newblock In {\em Proceedings of the 22nd international conference on World
  Wide Web}, 2013.

\bibitem{vallina2012breaking}
Narseo Vallina-Rodriguez, Jay Shah, Alessandro Finamore, Yan Grunenberger,
  Konstantina Papagiannaki, Hamed Haddadi, and Jon Crowcroft.
\newblock Breaking for commercials: characterizing mobile advertising.
\newblock In {\em Proceedings of the Internet Measurement Conference}, 2012.

\bibitem{pachilakis2019no}
Michalis Pachilakis, Panagiotis Papadopoulos, Evangelos~P Markatos, and Nicolas
  Kourtellis.
\newblock No more chasing waterfalls: A measurement study of the header bidding
  ad-ecosystem.
\newblock In {\em Proceedings of the 19th Internet Measurement Conference},
  2019.

\bibitem{gill2013best}
Phillipa Gill, Vijay Erramilli, Augustin Chaintreau, Balachander Krishnamurthy,
  Konstantina Papagiannaki, and Pablo Rodriguez.
\newblock Follow the money: understanding economics of online aggregation and
  advertising.
\newblock In {\em Proceedings of the Conference on Internet measurement
  conference}, 2013.

\bibitem{reznichenko2011auctions}
Alexey Reznichenko, Saikat Guha, and Paul Francis.
\newblock Auctions in do-not-track compliant internet advertising.
\newblock In {\em Proceedings of the 18th ACM conference on Computer and
  communications security}, 2011.

\bibitem{rtbPrices17}
Panagiotis Papadopoulos, Nicolas Kourtellis, Pablo~Rodriguez Rodriguez, and
  Nikolaos Laoutaris.
\newblock If you are not paying for it, you are the product: How much do
  advertisers pay to reach you?
\newblock In {\em Proceedings of the Internet Measurement Conference}, 2017.

\bibitem{haddadi2009not}
Hamed Haddadi, Saikat Guha, and Paul Francis.
\newblock Not all adware is badware: Towards privacy-aware advertising.
\newblock In {\em Conference on e-Business, e-Services and e-Society}, 2009.

\bibitem{reznichenko2014private}
Alexey Reznichenko and Paul Francis.
\newblock Private-by-design advertising meets the real world.
\newblock In {\em Proceedings of the ACM SIGSAC Conference on Computer and
  Communications Security}, 2014.

\bibitem{backes2012obliviad}
Michael Backes, Aniket Kate, Matteo Maffei, and Kim Pecina.
\newblock Obliviad: Provably secure and practical online behavioral
  advertising.
\newblock In {\em IEEE Symposium on Security and Privacy}, 2012.

\bibitem{10.1145/1859983.1859993}
Hamed Haddadi, Pan Hui, and Ian Brown.
\newblock Mobiad: Private and scalable mobile advertising.
\newblock In {\em Proceedings of the Fifth ACM International Workshop on
  Mobility in the Evolving Internet Architecture}, 2010.

\bibitem{biswas2014privacy}
Debmalya Biswas and Krishnamurthy Vidyasankar.
\newblock Privacy preserving and transactional advertising for mobile services.
\newblock {\em Computing}, 2014.

\bibitem{10.1145/2534169.2486038}
Sebastian Angel and Michael Walfish.
\newblock Verifiable auctions for online ad exchanges.
\newblock {\em SIGCOMM Comput. Commun. Rev.}, 2013.

\bibitem{pang2015prota}
Yiming Pang, Bo~Wang, Fan Wu, Guihai Chen, and Bo~Sheng.
\newblock Prota: A privacy-preserving protocol for real-time targeted
  advertising.
\newblock In {\em 2015 IEEE 34th International Performance Computing and
  Communications Conference (IPCCC)}. IEEE, 2015.

\bibitem{5958026}
M.~{Fredrikson} and B.~{Livshits}.
\newblock Repriv: Re-imagining content personalization and in-browser privacy.
\newblock In {\em 2011 IEEE Symposium on Security and Privacy}, SP'11, 2011.

\bibitem{8228673}
A.~{Mandal}, J.~{Mitchell}, H.~{Montgomery}, and A.~{Roy}.
\newblock Privacy for targeted advertising.
\newblock In {\em 2017 IEEE Conference on Communications and Network Security
  (CNS)}, 2017.

\bibitem{wang2015privacy}
Wei Wang, Linlin Yang, Yanjiao Chen, and Qian Zhang.
\newblock A privacy-aware framework for targeted advertising.
\newblock {\em Computer Networks}, 2015.

\bibitem{parra2017pay}
Javier Parra-Arnau.
\newblock Pay-per-tracking: A collaborative masking model for web browsing.
\newblock {\em Information Sciences}, 2017.

\end{thebibliography}
